\documentstyle[preprint,aps,psfig]{revtex}
\newcommand{\be}{\begin{equation}}
\newcommand{\ee}{\end{equation}}
\newcommand{\bea}{\begin{eqnarray}}
\newcommand{\eea}{\end{eqnarray}}
\newcommand{\ta}{\tilde\alpha}
\newcommand{\tb}{\tilde\beta}

\newcommand{\da}{\dot\alpha}
\newcommand{\db}{\dot\beta}

\tightenlines
\begin{document}
\tighten
\title{SO(10) A LA  PATI-SALAM }
\author{
Charanjit S. Aulakh{\footnote{Email: aulakh@pu.ac.in} and Aarti
Girdhar}
\\ {\it Dept. of Physics, Panjab University,Chandigarh, India}\\}
\maketitle
\begin{abstract}
We present rules for rewriting SO(10) tensor and spinor invariants
in terms of invariants of its ``Pati-Salam'' maximal subgroup
(SU(4)$\times \rm{SU(2)}_L\times \rm{SU(2)}_R )$ supplemented by
the discrete symmetry called D parity. Explicit decompositions of
quadratic and cubic invariants relevant to GUT model building are
presented and the role of D parity in organizing the terms
explained. Our rules provide a complete and explicit method for
obtaining the ``Clebsch-Gordon" Coefficients for $SO(10)\leftrightarrow G_{PS}$ in
a notation appropriate for  field theory models. We
illustrate the usefulness our methods by calculating previously unavailable 
mass matrices and couplings of the $SU(2)_L$ doublets and $SU(3)_c $ 
triplets in the  minimal Susy SO(10) GUT which are essential
to specify the   phenomenology of this model. We also present the 
bare effective potential for Baryon number violation in this model and
show that it recives novel contributions  from exchange of triplet Higgsinos  
contained the in ``neutrino  mass''
 Higgs submultiplets ${\bf{\overline{\Sigma}}}_{126}(10,1,3)$.
This further tightens the emerging connection between neutrino mass and proton decay.  
 
\end{abstract}
\section {Introduction}
The virtues of  SO(10) supersymmetric GUTs
\cite{dw}-\cite{so10refs}
 are now widely appreciated. SO(10) has the
cardinal virtue of exactly accommodating, within a single (16
dimensional) irrep, the 15 chiral fermions of a Standard Model
family plus the right handed neutrino, which now has a strong
claim to inclusion in any fundamental theory since neutrino masses
are an inalienable part of particle phenomenology \cite{sk,sno}. Thus the seesaw
mechanism \cite{seesaw,mohsen} finds a natural
home in SO(10). Moreover SO(10) provides an appealing rationale
for the parity breaking manifest in the  Standard model by linking
it to the breaking of Left-Right symmetry which embeds naturally
in SO(10) via its Pati-Salam\cite{ps74} maximal subgroup $
G_{PS}=\rm{SU(4)}\times \rm{SU(2)}_L\times \rm{SU(2)}_R $ ( More precisely
$G_{PS} \times D$, where D is the so called D
parity\cite{Dparity,cmp}).

There are, however, two contending points of view regarding the
type of  Higgs fields that should be used. Specifically, the
question is whether \cite{dw}-
\cite{abmrs01}, or not \cite{so10refs}, large tensor
representations like the 126 may be legitimately employed in view
of their strong effect on the SO(10) beta function above the GUT
scale and the difficulty of obtaining them from string theory. In
supersymmetric models of the first type (which employ a
``renormalizable see-saw mechanism'' based on even B-L Higgs
multiplets lying within the  $\bf\overline{126}$ Higgs) the
crucial R/M-parity of the MSSM becomes a part of
 the gauge symmetry and demonstrably survives symmetry
breaking \cite{abmrs01},\cite{abs}-
\cite{amrs99}. In the alternative viewpoint\cite{so10refs} the use of $SO(10)$
spinorial $\bf{16},{\bf{{\overline{16}}}}$ plet Higgs is advocated with nonrenormalizable
couplings providing the effective $\bf{\overline{126}}$
dimensional operators needed for giving a large Majorana mass to
the right handed neutrino. Other{\it{ ad hoc}} symmetries are
employed to play the role of R/M-parity which is strongly broken,
obliterating the distinction between Higgs and sfermion scalars in
the fundamental theory. This approach has the virtue of smaller
threshold effects at the GUT scale and moreover the theory does
not necessarily become asymptotically strong  very close to the
scale of perturbative Grand Unification. On the other hand it has
recently been argued \cite{trm,tas} that the explosion in the
gauge coupling constant just above the GUT scale, due to the
inclusion of  Higgs multiplets adequate to achieve realistic tree
level matter mass spectra,  is in fact the flag of a new type of
UV strong dynamical GUT symmetry breaking due to formation of SM
singlet condensates , which can be analysed (since $M_{GUT}=M_U
>> M_{Susy}=M_S$), using the methods  (based on holomorphy of
F-terms) developed by Seiberg and others \cite{seiberg} for
supersymmetric gauge theories. In either type of theory knowledge
of the Clebsch-Gordon coefficients for SO(10) or equivalently the
ability to break up SO(10) invariants into those of its subgroups
$G_{PS},\supset G_{LR},\supset G_{123}$ is essential.

In previous work\cite{aulmoh,abmrs01,ams,amrs,abmrs} it was shown
that in supersymmetric theories the restricted form of the
superpotential can leave Renormalization Group (RG) significant
multiplets with only intermediate or even light masses. Thus a
proper RG analysis of Susy GUTs should make use of the actual mass
spectrum of the model in question rather than the spectrum
conjectured on the basis of the survival principle. To implement
this program it is necessary to formulate the matching conditions
for the couplings of the various mass multiplets at successive
symmetry breaking and mass thresholds of the theory. Since the low
energy theory is based upon a unitary gauge group whereas the
ultimate determinant of coupling constant relations is the
overlying SO(10) gauge symmetry it is necessary to write the
SO(10) invariants in terms of properly normalized fields carrying
the unitary maximal subgroup labels. The initial work on the
minimal Susy GUT based on the 210-plet of SO(10) \cite{aulmoh,ckn}
was followed by an analysis of some of the SO(10) Clebsch-Gordon
coeffcients in \cite{heme,lee}, which, however, could  yield only
incomplete results. The maximal subgroups of SO(10) are
$SU(5)\times U(1)$ and the Pati-Salam Group $SU(4)\times
SU(2)_L\times SU(2)_R $ which is isomorphic to the  $SO(6)\times
SO(4)$ subgroup of $SO(10)$. Very recently \cite{nathraza} the
explicit forms of the SO(10) invariants of representations (with
dimensions upto 210) were given in terms of $SU(5)\times U(1)$
labels using the so called oscillator basis\cite{mohsak} to effect
the conversion. This rewriting, besides suffering from a certain
lack of transparency (due precisely to the LR asymmetric nature of
the embedding of $ SU(5)\times U(1)$),~is quite inappropriate for
LR symmetric breaking chains. Thus it is necessary to obtain the
invariants in terms of the PS subgroup separately. Moreover our
results may be reassembled into $SU(5)\times U(1)$ invariants and
can serve as an alternative derivation and cross check.

Furthermore, a discrete symmetry closely related to
Parity\cite{Dparity}, namely the so called D-parity, is important
and useful in studying the possible symmetry breaking chains in
SO(10) GUTs {\cite{abmrs01,cmp,geneal}}. In the decomposition of
SO(10) invariants into PS invariants D-parity proves valuable
for organizing and cross checking relative signs in our
expressions. We have developed explicit rules for the action of
D-parity on all fields according to their (SO(10) tensor or
spinor) origin and their PS labels.

Although the necessary basic tools have long existed (in somewhat
implicit form) in the work of Wilczek and Zee \cite{wilzee} no
explicit results are available. Moreover we disagree with
\cite{wilzee} regarding the explicit form of the possible Charge
conjugation matrices to be used for $SO(2N)$ spinorial
representations. Indeed it is only after making the necessary
corrections that the translation $SO(10) \leftrightarrow G_{PS}$
becomes feasible and transparent. Therefore we have attempted to
fill the long standing lacuna and provided rules for the
translation from SO(10) labels to the PS unitary subgroup labels.
Our results immediately allow us to derive the mass matrices of 
certain $SU(2)_L$ doublets and $SU(c)_c$ triplets which are crucial to 
specifying the low energy effective theory as the MSSM and to derive the 
bare effective potential describing the  most distinctive signature of GUTs 
namely Baryon violation. This unveils a new 
contribution to baryon decay mediated by colour triplets contained in the 
PS ``neutrino mass " decuplet-triplets : ${\bf{\overline{\Sigma}_{126}(10,1,3)}}$
and further strengthens the likely link between Baryon violation and neutrino mass that 
surfaced post-Super Kamiokande \cite{babpatwil}.  
The calculation is performed in the context of  the recently revived\cite{abmsv,fukujanac} ``minimal 
supersymmetric GUT" \cite{aulmoh,ckn,heme,lee}
based on the $\bf{210}$-plet Higgs  which was proposed more than 
20 years ago \cite{aulmoh,ckn} but still lacked the coupling
coefficients we have provided and which, to our knowledge , 
are not easily and explicitly obtainable by any other method.

In Section II we introduce our notation and the embedding of
SO(6)$\times$ SO(4) in SO(10) and define D-parity on tensor
representations. We then show how to rewrite invariants formed
from  SO(6) tensor irreps  in terms of SU(4) labels, and similarly
for SO(4) invariants to $\rm{SU(2)}_L\times \rm{SU(2)}_R$ labels.
In Section III we implement these rules on some tensor invariants
to illustrate the procedures for translating from SO(10) to
$G_{PS}$. However, since an  exhaustive listing of invariants is
both exhausting to produce and counterproductive as regards actual
utility for users of these techniques, we have instead provided an
Appendix where we collect useful SO(6) and SO(4) contractions
translated to unitary form. This collection permits easy
computation of SO(10) invariants formed from any tensor
representation of dimension $\leq$ 210. In Section IV, V we perform
the same tasks once spinor representations are included. In
Section VI we apply our results to compute the phenomenologically crucial
Electroweak doublet and  Colour triplet mass matrices 
in the minimal Susy SO(10) GUT of
\cite{aulmoh,ckn,lee,abmsv}. We also calculate 
explicitly the bare effective superpotential for Baryon number violation 
in this model .  We conclude with some remarks on
future directions.
\section{SO(10) $\rightarrow$ SO(6) $\times$ SO(4) $\sim$  $G_{PS}$}
The PS subgroup  SU(4) $\times \rm{SU(2)}_L \times \rm{SU(2)}_R$
$\subset$ SO(10) is actually isomorphic to the obvious maximal
subgroup SO(6) $\times$ SO(4) $\subset$ SO(10). The essential
components of the analysis are thus explicit translation between SO(6) and
SU(4) on the one hand and SO(4) and SU(2)$_L \times  \rm{SU(2)}_R$ on the
other.
Our notations and conventions follow those of \cite{wilzee} wherever
possible. Wherever feasible we repeat definitions so that the presentation is
self contained.  A crucial difference with \cite{wilzee} concerning the
explicit form of the charge conjugation matrices for spinor
representations of orthogonal groups will however emerge in the section
on spinors.

We have adopted the rule that any submultiplet of an SO(10) field is
always denoted by the {\it{same}} symbol as its parent field, its
identity being
established by the indices it carries or by supplementary indices, if
necessary. Our notation for indices is as follows : The indices of the
vector representation of SO(10) (sometimes also SO(2N))
 are denoted by $i,j =1..10 (2N)$. The {\it {real}}
vector index of the upper left block embedding (i.e. the embedding
specified by the breakup of the vector multiplet $10=6 + 4$) of
SO(6) in SO(10) are denoted $a,b=1,2..6$ and of the lower right
block embedding of SO(4) in SO(10) by
${\tilde{\alpha},\tilde\beta= 7,8,9,10}$. These indices are
complexified via a Unitary transformation and denoted by
$\hat{a},\hat{b}=\hat{1},\hat{2},\hat{3},\hat{4},\hat{5},\hat{6}
\equiv \overline{\mu},\overline{\mu}^{*}=
\bar{1},{\bar{1}^*,\bar{2},\bar{2}^*,\bar{3},\bar{3}^*}$ where
$\hat{1}\equiv \bar{1}, \hat{2}\equiv \bar{1}^* $ etc. Similarly
we denote the complexified versions of
${\tilde{\alpha},\tilde\beta}$ by $\hat{\alpha},\hat{\beta}=
\hat{7},\hat{8},\hat{9},\widehat{10}$. The indices of the doublet
of SU(2)$_L(\rm{SU(2)}_R$) are denoted
$\alpha,\beta=1,2$($\dot\alpha,\dot\beta=\dot{1},\dot{2})$.
Finally the index of the fundamental 4-plet of SU(4) is denoted
by a (lower) $\mu,\nu = 1,2,3,4$ and its upper-left block SU(3)
subgroup indices are $\bar\mu,\bar\nu = 1,2,3$. The corresponding
indices on the $4^{*}$ are carried as superscripts.
 
\subsection  {\ SO(6) $ \longleftrightarrow  $\ SU(4)}
{{\bf{\underline{Vector/Antisymmetric}}}:  \vspace{.1 cm} \noindent
The 6 dimensional vector representation of SO(6) denoted by $V_a
 (a=1,2,..,6)$ transforms as
\begin{equation}
V'_a=(exp {i \over 2}\omega ^{cd}J_{cd})_{ab}V_b
\end{equation}
where the Hermitian generators $J_{cd}$ have the explicit form
\begin{equation}
(J_{cd})_{ef}=-i\delta_{c[e}\delta_{f]d}
\end{equation}
and thus satisfy the SO(6) algebra (square brackets around indices  denote antisymmetrization)
\begin{equation}
[J_{cd},J_{ef}]= i\delta_{e[c}J_{d]f}-i\delta_{f[c}J_{d]e}
\end{equation}
It is useful to introduce complex indices
$\hat a, \hat b = {\hat 1} ...{\hat 6}  $ by the unitary change of
basis
\bea
V_{\hat a}=U_{\hat a {a}}V_a \;, \quad U=U_2 \times I_3\;, \quad U_2={1
\over \sqrt{2}}
   \left[  \begin{array}{cc}
1 & i \\ 1 & -i
\end{array}\right]
\eea
so that  $V_{a}W_{a}=V_{\hat a}W_{\hat a^*}$. The decomposition of the
fundamental 4-plet of  SU(4) w.r.t. SU(3)$\times \rm{U(1)}_{B-L}$ is
$4=(3,1/3) \oplus(1,-1)$. The index for the 4 of SU(4) is denoted by
$\mu=1,2,3,4$
while $\bar \mu=1,2,3$ label its SU(3) subgroup.
In SU(4) labels,~the 6 of  SO(6) is the 2 index antisymmetric
$V_{\mu\nu}$ and decomposes as $6=V_{\bar\mu}(3,-2/3) \oplus V_{{\bar\mu}^*
}(\bar 3, 2/3)$ and
we identify $V_{\bar\mu4}=V_{\bar\mu}$,~$V_{\bar\mu\bar
\nu}=\epsilon_{\bar\mu \bar \nu\bar \lambda}V_{\bar\lambda^*}$. In other
words, if one defines $V_{\mu\nu}=\Theta_{\mu\nu}^{\hat a} V_{\hat a}$ with
$\Theta_{{\bar \mu} 4}^{\hat a}=\delta^{\hat a}_{\bar \mu},
 \Theta_{{\bar \mu}{\bar \nu}}^{\hat a} =
\epsilon_{\bar \mu \bar \nu \bar \lambda}\delta^{\hat a}_{\bar
\lambda^*}$,~then since $\Theta_{\mu\nu}^{\hat a} \Theta_{\lambda\sigma}^
{\hat a*} \equiv \epsilon_{\mu\nu\lambda\sigma}$ it follows that the
translation of SO(6) vector index contraction is (${\widetilde V}^{\mu\nu}
=(1/2)\epsilon^{\mu\nu\lambda\sigma} V_{\lambda \sigma})$
\bea
V_aW_a&=&{1
\over4}\epsilon^{\mu\nu\lambda\sigma}V_{\mu\nu}W_{\lambda\sigma}
\equiv {1 \over2}\widetilde V^{\mu\nu}W_{\mu\nu}\label{6vec}\\
{\rm{while}}~~~
V_aW_a^*&=&{1 \over 2} V_{\mu\nu}W_{\mu\nu}^*
\eea
Representations carrying vector indices $a,b  ...$ are then translated
by
replacing by each vector index by an antisymmetrized pair of
SU(4) indices $\mu_1\nu_1, \mu_2\nu_2......$. For example
\bea
A_{ab} B_{ab} &= &2^{-4} \epsilon^{\mu_1\mu_2\mu_3\mu_4}
\epsilon^{\nu_1\nu_2\nu_3\nu_4} A_{{\mu_1\mu_2},{\nu_1\nu_2}}
B_{{\mu_3\mu_4},{\nu_3\nu_4}}\\
{\rm{while}}~~~~~
A_{ab} B^*_{ab}& =& 2^{-2}  A_{{\mu_1\mu_2},{\nu_1\nu_2}}
B^*_{{\mu_1\mu_2},{\nu_1\nu_2}}
\eea
{\bf{\underline{Antisymmetric/Adjoint}}}:
\vspace{.1 cm}\noindent
The 15 dimensional antisymmetric representation $A_{ab}$ of SO(6) translates to the adjoint
 {\bf{15}} $A{_{\nu}}
^{\mu}$ of SU(4):
\be
A{{_\nu}}^{\mu}=+{1 \over
4}\epsilon^{\mu\lambda\sigma\rho}A_{\lambda\sigma,\rho\nu}=
-A^{\mu}_{~\nu}\quad  ;\quad
A_{\mu\nu,\rho\sigma}=+\epsilon_{\lambda\mu\nu[\rho}A_{\sigma]}^{~~\lambda}
\label{AD} \ee The parameters $\omega_{ab}$ of SO(6) are
identified with those of SU(4) $({\theta^A, A=1...15})$
\be
\omega_{ab} \rightarrow {\omega_{\mu}}^{\nu} = i\theta^{A}{(
\lambda^A)_{\mu}}^{\nu} \label{param}\ee Where $\lambda^A, A=1..15 $ are the
Gellmann matrices for $SU(4)$ and the group element in in the
fundamental is $exp({i\theta^{A}\lambda^{A}\over{2}})$. We define
\bea {A_{\nu}}^{\mu}={i \over
{\sqrt2}}{(\lambda^A)}_{\nu}^{~\mu}A^A~,\qquad
{(\lambda^A)}_{\nu}^{~~\mu} \equiv {\lambda^A}_{\nu\mu} \eea
Note that tracelessness $A{{_\mu}^\mu}=0$ is ensured by antisymmetry of
$A_{\mu\nu,\lambda\sigma}$ and symmetry of
$\epsilon^{\mu\nu\lambda\sigma}$ under interchange of index pairs $\mu\nu$
and $\lambda\sigma$. The normalization relation
\bea
(A{{_\nu}^\mu},A{{_\sigma}^\lambda})&=&
\delta^{\lambda}_{\mu}\delta^{\nu}_
{\sigma}-{1 \over 4}\delta^{\nu}_{\mu}\delta^{\lambda}_{\sigma}
\nonumber\\&=&
{1 \over 2}((\lambda^A){{_\nu}^\mu})^*(\lambda^A){{_\sigma}^\lambda}
\eea
follows if
 $A_{ab}, A^A$ are of unit norm :
 \be
(A_{ab}, A_{cd})=\delta_{a[c}\delta_{d]b}~~ ;~~(A^{A}, A^{B})=\delta^{AB}
\ee
We denote the trace over SO(6) vector indices a,b ... by ``Tr'' and
over the SU(4) fundamental index $\mu\nu...$ by ``tr''.
 Then
\bea
TrAB &=& A_{ab} B_{ba} = 2 A{{_\nu}^\mu} B{{_\mu}^\nu}=2 trAB
 \nonumber\\
TrABC &=& -trA [B,C] \eea A notable point is that the invariant 6
index totally antisymmetric tensor of SO(6) leads to a distinct
$SU(4)$ invariant involving the anti- commutator.
\be
\epsilon_{abcdef}A_{ab} B_{cd} C_{ef} = -8i(trA {\{B,C\}}) \ee
{\bf{\underline{Symmetric traceless (20)/4 index mixed}}}:
\vspace{.1 cm}\noindent
The 20 dimensional symmetric traceless representation
${ S}_{ab}$ of
SO(6) which has normalization
\be ({ S}_{ab}, { S}_{cd}) = \delta^a_{(c} \delta^b_{d)} -
{1\over 3}\delta^{ab} \delta_{cd}
\ee
appropriate to a traceless field
translates to  ${ S}_{\mu\nu,\lambda\sigma}={ S}_{\lambda\sigma,\mu\nu}$ with
the
additional constraint (corresponding to tracelessness on SO(6) vector
indices)
\be
{1\over 4}\epsilon^{\mu\nu\lambda\sigma}{ S}_{\mu\nu,\lambda\sigma}
\equiv { S}_{aa}=0
\ee
The normalization condition translates to
\be
({ S}_{\mu\nu,\lambda\sigma},{ S}_{\theta\delta,\epsilon\rho}) =
\delta^{\mu}_{[\theta} \delta^{\nu}_{\delta]}
\delta^{\lambda}_{[\epsilon}
\delta^{\sigma}_{\rho]} + \delta^{\mu}_{[\epsilon}\delta^{\nu}_{\rho]}
\delta^{\lambda}_{[\theta}\delta^{\sigma}_{\delta]} -{1\over 3}
\epsilon^{\mu\nu\lambda\sigma} \epsilon_{\theta\delta\epsilon\rho}
\ee
{\bf{\underline{3 Index Antisymmetric (Anti) Self Dual/Symmetric 2 index}}}:
\vspace{.1cm}\noindent
The invariant tensor $\epsilon_{abcdef}$ of SO(6) allows
the separation of the 3 index totally antisymmetric 20-plet
$T_{abc}$ of SO(6) into self dual and anti-self dual pieces
$T^\pm_{abc}=\pm
\widetilde T^\pm_{abc}$ where the SO(6) dual is defined as
\be
\widetilde T_{abc}={i \over 3!}\epsilon_{abcdef}T_{def}\label{tT}
\ee
$T^{+}_{abc}(T^{-}_{abc})$ translate into the 2 index symmetric
$10 (T_{\mu\nu})~~(\overline{10}({\overline T}^{\mu\nu}))$ of SU(4) via
\bea
T_{\mu\nu}&=&{1 \over 12}T^{+}_{\mu\lambda,\nu\sigma,\gamma\delta}
\epsilon^{\lambda\sigma\gamma\delta}\label{sd}\\
\overline T^{\mu\nu}&=&{1 \over 24}T^-_{\kappa\lambda,\rho\sigma,\pi\theta}
\epsilon^{\mu\kappa\lambda\pi}\epsilon^{\nu\rho\sigma\theta}\label{asd}\\
T^{(+)}_{\mu\nu,\rho\theta,\gamma\delta}&=&
T_{{_[\mu}{^[\rho}}\epsilon_{{\nu_]}
{\theta^]}\gamma\delta}\label{TPD}\\
T^{(-)}_{\kappa\lambda,\theta\rho,\sigma\delta}&=&-{\overline T}^{\mu\nu}
\epsilon_{\mu\kappa\lambda_[\sigma}\epsilon_{\delta_]\nu\theta\rho}\label{TMD}
\eea
Note that to preserve unit norm one should define
\be
T^\pm_{abc}={{{T_{abc}\pm\widetilde T_{abc}} \over\sqrt2}}\label{tp}
\ee
The normalization conditions that follow from unit norm for
$T_{abc}$ :
\be
{(T_{abc},T_{a'b'c'})}={\delta^a_{[a'} \delta^b_{b'}
\delta^c_{c']}}\label{tn1}\\
\ee
are
\be
{(T_{\mu\nu},T_{\lambda\sigma})}=\delta^{\mu}_{(\lambda}\delta^{\nu}_{\sigma)}
={(\overline T^{\lambda\sigma},\overline T^{\mu\nu})}\label{tn2}
\ee
So that $T_{\mu\mu}$(no sum) has norm squared 2 while
$T_{\mu\nu}(\mu\neq \nu)$ has norm one.\\
One has the useful identity~:~
$T^{+}_{abc}T^{-}_{abc}=6~T_{\mu\nu}~\overline{T}^{\mu\nu}$
\subsection{$\rm{SO(4)} \leftrightarrow \rm{SU(2)}_{L} \times \rm{SU(2)}_{R}$}
{{\bf{\underline{Vector/Bidoublet}}} \vspace{.1 cm}\\ We use early
greek indices ${ \ta, \tb =7,8,9,10}$ for the vector of SO(4)
corresponding to ${i,j=7....10}$ of the 10-plet of SO(10). The
Hermitian generators of SO(4) have the usual SO(2N) vector
representation form :
${(J_{\tilde\alpha\tilde\beta})_{\tilde\gamma\tilde
\delta}}=-i{\delta_{\tilde\alpha[\tilde\gamma}\delta_{\tilde\delta]\tilde
\beta}} $.\\ The group element is $R=exp {i \over
2}\omega^{\tilde\alpha\tilde\beta}J_ {\tilde\alpha\tilde\beta}$.
The generators of SO(4) separate neatly into self-dual and
anti-self- dual sets of 3, ${J_{\tilde
\alpha\tilde\beta}^{\pm}}={1 \over 2}{(J_{\tilde\alpha\tilde\beta}
\pm {\tilde J_{\tilde\alpha\tilde\beta})}}$. Then if
${\check\alpha,\check\beta=1,2,3}$ the generators and parameters
of the $SU(2)_{\pm}$ subgroups of SO(4) are defined to be
\be
J_{\check\alpha}^{\pm}={1 \over
2}\epsilon_{\check\alpha\check\beta\check \gamma} J_{(\check\beta
+6)(\check\gamma +6)}^{\pm} ~ ;\quad
\omega_{\check\alpha}^{\pm}={1 \over 2}\epsilon_{{\check
\alpha}{\check \beta }{\check\gamma}}\omega_{{(\check\beta+6})
(\check\gamma +6) } \pm \omega_{(\check\alpha+6)10} \ee
The $SU(2)_{\pm}$ group elements are ${exp(i{\vec \omega}_{\pm} \cdot \vec J^{\pm})}.$
The vector 4-plet of SO(4) is a bi-doublet $(2,2)$ w.r.t. to
${SU(2)_{-} \otimes SU(2)_{+}}$.
We denote the indices of the doublet of ${SU(2)_L=SU(2)_{-}}$
${(SU(2)_R= SU(2)_{+})}$ by undotted early greek indices
${\alpha,\beta=1,2}$ {(dotted early greek indices
$\dot\alpha,\dot\beta=\dot{1},\dot{2}$)}. Then one has \bea
V_{\hat7}&=& V_{\bar 4}= {{(V_7+iV_8)} \over \sqrt2}=V_{2\dot2} ~,~ V_{\hat9}= V_{\bar 5}=
{{(V_9+iV_{10})}\over \sqrt2}=V_{1\dot2}\\
V_{\hat 8}&=& V_{\bar4^*}={{(V_7-iV_8)} \over \sqrt2}=-V_{1\dot1}~,
 ~V_{\widehat{10}}\equiv V_{\hat 0}= V_{\bar5^*}={{(V_9-iV_{10})}\over \sqrt2}=V_{2\dot1}
\label{bidoub} \eea ${SU(2)_L {(SU(2)_R)}}$ indices are raised and
lowered with ${\epsilon^ {\alpha\beta},\epsilon_{\alpha\beta}}$
${(\epsilon^{\dot\alpha\dot\beta},
\epsilon_{\dot\alpha\dot\beta})}$ with
${\epsilon^{12}=+\epsilon_{21}=1}$ etc. The SO(4) vector index
contraction translates as \bea V_{\tilde\alpha}W_{\tilde\alpha}
&=& -{V_{\alpha\dot\alpha}W_{\beta\dot\beta}
\epsilon^{\alpha\beta}\epsilon^{\dot\alpha\dot\beta}}
=-{V^{\alpha\dot\alpha}W_{\alpha\dot\alpha}}\\ {\rm{While}}~~~~~~~
V_{\tilde\alpha}W_{\tilde\alpha}^*
&=&{V_{\alpha\dot\alpha}W^{*}_{\alpha\dot\alpha}}\label{4vec} \eea
{\bf{\underline{Antisymmetric Selfdual/triplet}}}~:~\noindent
Separating the 2 index antisymmmetric tensor
$A_{\tilde\alpha\tilde\beta}$ into self-dual and anti-self-dual
parts of unit norm
\be
A^{(\pm)}_{\ta\tb}={1 \over \sqrt2}{(A_{\ta\tb} \pm\tilde{A}_{\ta\tb})}\label
{at}
\ee
One finds ${A^{-}{(A^{+})}}$ is ${{(3,1)}{((1,3))}}$ w.r.t.
${SU(2)_L \times SU(2)_R}$. In fact these triplets are just \bea
A_{\check\alpha}^{(\pm)}&=& {\pm} A^{(\pm)}_{{\check\alpha +6},10}
\nonumber
\\&=& {1 \over 2}\epsilon
_{\check\alpha \check\beta \check\gamma }
A_{(\check\beta+6)(\check\gamma +6)}^{(\pm)}
\eea
Defining ${{A_\alpha}^\beta}=i
A_{\check \alpha}^{(-)}{{(\sigma^{\check\alpha})_\alpha}^\beta}=i\vec{A}_{L} \cdot
(\vec\sigma)_{\alpha}^{~~\beta}$~~,
 ${{A_{\dot\alpha}}^{\dot\beta}}=i A_{\check\alpha}^{(+)}{(\sigma^{\check
 \alpha})
 _{\dot\alpha}}^{\dot\beta}=i\vec{A}^{R} \cdot (\vec\sigma)_{\dot\alpha}^
 {~\dot\beta}$, where $\sigma^{\check\alpha}$ are the Pauli matrices,
one has
\bea
A_{\hat\alpha\hat\beta}^{(+)}& \rightarrow&  A_{\alpha\dot\alpha\beta\dot
\beta}^{(+)}\equiv\epsilon_{\alpha\beta}A_{\dot\alpha\dot\beta}=\epsilon_
{\alpha\beta}A_{\dot\beta\dot\alpha}\label{ap}\\
A_{\hat\alpha\hat\beta}^{(-)} &\rightarrow& A_{\alpha\dot\alpha\beta\dot\beta}
^{(-)}\equiv\epsilon_{\dot\alpha\dot\beta}A_{\alpha\beta}=\epsilon_{\dot
\alpha\dot\beta}A_{\beta\alpha}\label{am}
\eea
Where the index pairs $\alpha\dot\alpha$ correspond to the complex
indices
 $\hat\alpha$ as given in (\ref{bidoub}) above.
Then one has for the contraction of two antisymmetric tensors
\bea
A_{\ta\tb}{B}_{\ta\tb}&=&{1 \over
2}{(A_{\ta\tb}^{(+)}{B}_{\ta\tb}^{(+)}+
 A_{\ta\tb}^{(-)}{B}_{\ta\tb}^{(-)})}\\
 &=& 2{(\vec{A}_L\cdot\vec{B}_L+\vec{A}_R\cdot\vec{B}_R)}
 \eea
Similarly one gets the useful identity
\be
A^{(\pm)}_{\ta\tb}B^{(\pm)}_{\tb\tilde\gamma}C^{(\pm)}_{{\tilde\gamma}\ta}=
4 \vec{A}^{(\pm)} \cdot ({\vec{B}^{(\pm)} \times \vec{C}^{(\pm)}})
\ee
\vfil\eject
{\bf{\underline{Symmetric Traceless(9)/Bitriplet(3,3)}}}~:~
\vspace{.1cm}\noindent
The two index symmetric traceless tensor $S_{\ta\tb}$ of SO(4)
which has dimension
9 becomes the ${(3,3)}$ w.r.t $\rm{SU(2)}_L\times \rm{SU(2)}_R$ {(symmetry follows from tracelessness)}:
 \be
S_{\hat\alpha\hat\beta}=S_{\alpha\dot\alpha,\beta\dot\beta} \equiv S_
{\alpha\beta,
\dot\alpha\dot\beta}=S_{\beta\alpha,\dot\alpha\dot\beta}=S_{\alpha\beta,\dot
 \beta\dot\alpha}
 \ee
 so that e.g.
 \be
S_{\ta\tb}S'_{\ta\tb}=S^{\alpha\beta,\dot\alpha\dot\beta}S'_{\alpha\beta,\dot
 \alpha\dot\beta}
 \ee
 and are normalized as
 \be
{(S_{\alpha\beta,\dot\alpha\dot\beta},S_{\alpha'\beta',\dot\alpha'\dot\beta'}
)}=\delta^{\alpha}_{\alpha'}\delta^{\beta}_{\beta'}
\delta^{\dot\alpha}_{\dot\alpha'}
\delta^{\dot\beta}_{\dot\beta'}+\delta^{\alpha}_{\beta'}
\delta^{\beta}_{\alpha'}\delta^{\dot\alpha}_{\dot\beta'}
\delta^{\dot\beta}_{\dot\alpha'}-
{1 \over 2}\epsilon^{\alpha\beta} \epsilon^{\dot\alpha\dot\beta}
\epsilon_{\alpha'\beta'}\epsilon_{\dot\alpha'\dot\beta'}
 \ee
%
%
%
{\bf{\underline{SO(10) Tensors {\&} D-Parity}}} \vspace{.2 cm}\\
 The above treatment covers the
the SO(6) and SO(4) tensor representations encountered in dealing
with SO(10) representations upto dimension 210. The procedure for
the decomposition of SO(10) tensor invariants is now clear.
Splitting the summation over each SO(10) index i,j= 1,..10 into
summation over ${SO(6),SO(4)}$ indices $(a,\alpha)$, one replaces
each SO(6) (SO(4)) index by $SU(4)(SU(2)_L \times SU(2)_R)$ index
pair contractions according to the basic rules (\ref{6vec}) and
(\ref{4vec}) and uses (\ref{AD})(\ref{sd})(\ref{asd})(\ref{tp})
and (\ref{at})(\ref{ap})(\ref{am}) etc. to tranlate to PS labelled
fields and invariants.\\ An important and useful feature of the
decomposition is that it permits the transparent implementation of
the Discrete symmetry called D-Parity \cite{Dparity,cmp} defined
as
\be
D=exp(-i{\pi}J_{23})exp(i{\pi}J_{67})
\ee
On vectors this corresponds to rotations through  $\pi$ in the (23) and   (67)
planes. Thus components$(V_2,V_3,V_6,V_7)$ of $V_i$ change sign and the rest do not.
In PS language this becomes
\be
V_{\mu\nu} \leftrightarrow  (-)^{\mu+\nu+1}\widetilde{V}^{\mu\nu} \quad ,
\quad V_{2\dot{2}} \leftrightarrow V_{1\dot{1}}
\label{tendp1}\ee
While $V_{1\dot{2}},V_{2\dot{1}}$ remain unchanged. 
If we denote $\bar{1}=2$ and $\bar{2}=1$ for dotted and undotted
indices then these rules are just 
$V_{\alpha\dot\beta} \leftrightarrow V_{\dot{\bar\beta}
\dot{\bar\alpha}}$.

For the self-dual multiplets of SO(4) one finds that under D parity
\be
V^{(\pm)}_{1} \leftrightarrow V^{(\mp)}_{1} \quad ;\quad V^{(\pm)}_{2,3}
\leftrightarrow -V^{(\mp)}_{2,3}\\
\label{tendp2}\ee
I.e $V^{(-)}_{\alpha\beta} \leftrightarrow -V^{(+)}_{\dot{\bar\alpha}\dot
{\bar\beta}}$.
Then it follows that $\vec{A}_{L} \cdot \vec{B}_{L} \leftrightarrow
\vec{A}_{R} \cdot \vec{B}_{R}.$\\
The adjoint ${A_{\nu}}^{\mu}$ derived from the antisymmetric 15 has
D-parity property
\be
{A_{\nu}}^{\mu} \leftrightarrow (-)^{\mu+\nu+1}{A_{\mu}}^{\nu} \ee
On the other hand an adjoint derived from a 4 index antisymmetric
representation via \be \Phi_{ab}={1 \over {4!}}\epsilon_{abcdef}
\Phi_{cdef} \ee as occurs, for example, for (15,1,1) $\subset$ 210
and (15,2,2) $\subset$ 126, $\overline{126}$, will contain an
extra minus factor relative to (15,1,1) $\subset$ {\bf{45}}.
$\phi_{\nu}^{~\mu}\leftrightarrow (-)^{\mu+\nu}\phi_{\mu}^{~\nu}$
i.e. it is D-axial.\\ While the $SU(4)$ symmetric 10-plets from
the SO(6) (anti)self-dual 3 index antisymmetric transform as
\be
T_{\mu\nu} \leftrightarrow \overline{T}^{\mu\nu}(-)^{\mu+\nu+1}
\ee
\section {SO(10) Tensor Quadratic \& Cubic Invariants}
Using our rules we present examples of decompositions of SO(10) invariants
to illustrate the application of our method. As noted above, however, the
reader may find the generative rules collected
in the Appendix more convenient and complete in practice.\\
{\bf{\underline{$45 \cdot 45$}}}
\bea
45(A_{ij})&=&(15,1,1)A_{ab}+((1,3,1)A^{(-)}_{\ta\tb} \oplus
(1,1,3) A^{(+)}_{\ta\tb})+(6,2,2)A_{a\ta}\\
A_{ij}B_{ij}&=& A_{ab}B_{ab}+2A_{a\ta}
B_{a\ta}+A_{\ta\tb}B_{\ta\tb}\nonumber\\
&=&-2{A_{\nu}}^{\mu}{B_{\mu}}^{\nu}-A_{~~\mu\nu}^{\alpha\dot\alpha}{B}^
{\mu\nu}_{~~\alpha\dot\alpha}+2(\vec{A}_{L}.\vec{B}_{L}+\vec{A}_{R}.\vec{B}
_{R})
\eea
{\bf{\underline{$54 \cdot 54$}}}
\bea
54(S_{ij})&=&(20,1,1,)\widehat{S}_{ab}+(1,3,3)\widehat{S}_{\ta\tb}+
(6,2,2)S_{a\ta}+(1,1,1)S\\
S_{ij}R_{ij}&=&\widehat{S}_{ab}\widehat{R}_{ab}+\widehat{S}
_{\ta\tb}\widehat{R}_{\ta\tb}+2S_{a\ta}R_{a\ta}+2S.R\\
&=&{1 \over
4}\widehat{S}^{\mu\nu,\lambda\sigma}\widehat{R}_{\mu\nu,\lambda\sigma}+
\widehat{S}^{\alpha\beta,\dot\alpha\dot\beta}\widehat{R}_{\alpha\beta,\dot\alpha\dot\beta}-
{S}^{\mu\nu,\alpha\dot\alpha}{R}_{\mu\nu,\alpha\dot\alpha}+2S.R
\eea
\bea
{\rm{where}}~~~~ \widehat{S}_{ab}&=&{S}_{ab}-{\sqrt{2
\over 15}}\delta_{ab}S\\
\widehat{S}_{\ta\tb}&=&{S}_{\ta\tb}+{\sqrt{3 \over
10}}\delta_{\ta\tb}S\\ S&=&{\sqrt{5 \over 24}}S_{aa} \eea
{\bf{\underline{$54 \cdot 54 \cdot 54$}}}
\bea
S_{ij}R_{jk}T_{ki}&=&{1 \over {{2}^{3}}}\widehat{S}^{\mu\nu,\lambda\sigma}
{\widehat{R}_{\lambda\sigma}}^{~\theta\delta}{\widehat{T}_{\theta\delta,\mu\nu}}\nonumber\\
&-&\widehat{S}^{\alpha\beta,\dot\alpha\dot\beta}\widehat{R}_{\beta\gamma,
\dot\beta\dot\gamma}{\widehat{T}^{\gamma}_{~\alpha}},^{\dot\gamma}_{~\dot\alpha}
\nonumber\\
&-&{\sqrt {2 \over 15}}S.R.T \nonumber\\
&+&{\sqrt {1 \over 120}}{\{S}\widehat{R}^{\mu\nu,\lambda\sigma}\widehat{T}_
{\lambda\sigma,\mu\nu}+R{\widehat{S}}^{\mu\nu,\lambda\sigma}\widehat{T}_
{\lambda\sigma,\mu\nu}+T{\widehat{S}}^{\mu\nu,\lambda\sigma}\widehat{R}_
{\lambda\sigma,\mu\nu}\}\nonumber\\
&-&{{1 \over 4}}{\{\widehat{S}}_{\mu\nu,\lambda\sigma}{R^{\lambda\sigma}}_
{\alpha\dot\alpha}T^{\alpha\dot\alpha,\mu\nu}+\widehat{R}_{\mu\nu,\lambda
\sigma}{T^{\lambda\sigma}}_{\alpha\dot\alpha}S^{\alpha\dot\alpha,\mu\nu}+
\widehat{T}_{\mu\nu,\lambda\sigma}{S^{\lambda
\sigma}}_{\alpha\dot\alpha}R^{\alpha\dot
\alpha,\mu\nu}\}\nonumber\\
&-&{\sqrt {1 \over 120}}{\{S{R}^{\mu\nu,\alpha\dot\alpha}T_{\mu\nu,\alpha\dot
\alpha}+R{T}^{\mu\nu,\alpha\dot\alpha}S_{\mu\nu,\alpha\dot\alpha}+T{S}^{\mu\nu
,\alpha\dot\alpha}R_{\mu\nu,\alpha\dot\alpha}\}}\nonumber\\
&+&{1 \over 2}{\{\widehat{S}^{\alpha\beta,\dot\alpha\dot\beta}{R_{\beta\dot
\beta}}^{\mu\nu}T_{\mu\nu,\alpha\dot\alpha}+{\widehat{R}^{\alpha\beta,\dot
\alpha\dot\beta}}{T_{\beta\dot\beta}}^{\mu\nu}S_{\mu\nu,\alpha\dot\alpha}+
\widehat{T}^{\alpha\beta,\dot\alpha\dot\beta}{S_{\beta\dot\beta}}^{\mu\nu}
R_{\mu\nu,\alpha\dot\alpha}\}}\nonumber\\
&-&{\sqrt{3 \over 10}}{\{S\widehat{R}^{\alpha\beta,\dot\alpha\dot\beta}\widehat{T}
_{\alpha\beta,\dot\alpha\dot\beta}+R\widehat{T}^{\alpha\beta,\dot\alpha\dot\beta}
\widehat{S}_{\alpha\beta,\dot\alpha\dot\beta}+T\widehat{S}_{\alpha\beta,\dot\alpha
\dot\beta}\widehat{R}_{\alpha\beta,\dot\alpha\dot\beta}\}}
\eea
\vfil\eject
{\bf{\underline{$(45)^{2} \cdot 54 $}}}
\bea
A_{ij}A_{jk}S_{ki}&=&2{A_\lambda}^\mu{A_\sigma}^{\nu}{\widehat{S}_{\mu\nu}}
^{~\lambda\sigma}+{\sqrt{8 \over 15}}{A_\nu}^{\mu}{A_\mu}^{\nu}S
\nonumber\\
&+&{\sqrt{1 \over 30}}A^{\mu\nu,\alpha\dot\alpha}A_{\mu\nu,
\alpha\dot\alpha}S+{1 \over 4}{A_{\mu\nu}}^{\alpha\dot\alpha}A_
{\lambda\sigma,\alpha\dot\alpha}\widehat{S}^{\mu\nu,\lambda\sigma}\nonumber\\
&+&2 A^{\mu\nu,\alpha\dot\alpha}S_{\alpha\dot\alpha,\lambda\mu}{A_{\nu}}
^{\lambda}+{\sqrt{1 \over 2}}A^{\mu\nu,\beta\dot\beta}
(\epsilon_{\beta\alpha}A_{\dot\beta\dot\alpha}+\epsilon_{\dot\beta\dot\alpha}
A_{\beta\alpha}){S^{\alpha\dot\alpha}}_{\mu\nu}\nonumber\\
&-&{\sqrt{3 \over 40}}S{A}^{\mu\nu,\alpha\dot\alpha}A_{\mu\nu,\alpha\dot\alpha}
-{1 \over 2}\widehat{S}^{\alpha\beta,\dot\alpha\dot\beta}{A^{\mu\nu}}_{\alpha
\dot\alpha}A_{\mu\nu,\beta\dot\beta}\nonumber\\
&+&{\sqrt{6 \over 5}}S(\vec{A}_L.\vec{A}_L+\vec{A}_R.\vec{A}_R)-2A^{\dot
\alpha\dot\beta}A^{\alpha\beta}\widehat{S}_{\beta\alpha,\dot\beta\dot\alpha}
\eea
{\bf{\underline{$ \overline{126} \cdot 126$}}}
\bea
{1 \over 5!}\Sigma^{(-)}_{i_1...i_5}\Sigma^{(+)}_{i_1...i_5}
&=&\{\widetilde\Sigma^{(-){\mu\nu}}\Sigma^{(+)}_{\mu\nu}+2{\Sigma^
{(-)~~\mu}_{\nu}}^{\alpha\dot\alpha}\Sigma^{(+)\nu}_{\mu\alpha \dot\alpha}
\nonumber \\
& &+{\vec\Sigma}^{(-)}_{{R}\mu \nu} \cdot {\vec{\Sigma}^{(+)\mu \nu}}_{R}+
\vec{\Sigma}_{{L}\mu \nu}^{(+)} \cdot {\vec\Sigma}^{(-)\mu\nu}_{L}\}
\eea
Here $\Sigma^{(+)}({\bf{126}})
(\Sigma^{(-)}({\bf{\overline{126}}}))$ is the self-dual
 (antiself-dual) 5 index totally antisymmetric representation and 
the dual is defined as (note the minus sign)
\be
\widetilde\Sigma_{i_1...i_5}=-{i \over
5!}\epsilon_{i_1....i_{10}}\Sigma_ {i_6...i_{10}}
;~\widetilde\Sigma^{(\pm)}={\pm}\Sigma^{(\pm)}
\ee The SO(10) duality implies a correlation between the SO(6) and
SO(4) dualities of the SU(4) decuplet $\rm{SU(2)}_{L} \times \rm{SU(2)}_{R}$
triplets : \be +=(-,+)\oplus(+,-)~~,~~ -=(+,+)\oplus(-,-) \ee Where
$(-,+)$ refers to $(\overline{10},1,3)$ and $(+,-)$ to $(10,3,1)$.
So that, for example, $\Sigma^{+}$ has the decomposition \bea
\Sigma^{+}(126)&=&{\Sigma_{\nu}^{(+)\mu}}_{\alpha\dot\alpha}(15,2,2)+\vec
\Sigma^{(+)}_{\mu\nu~L}(10,3,1)\nonumber\\
&+&\vec\Sigma^{(+)~\mu\nu}_R(\overline{10},1,3) +\Sigma^{(+)}_{\mu\nu}(6,1,1)
\eea
While the $\Sigma^{-} (\overline{126})$ has the conjugate expansion.\\
\vspace{.1cm}
{\bf{\underline{$45 \cdot \overline{126} \cdot 126 $}}} ~:~ An example of the non trivial
action of D parity is given by
the terms containing the (15,1,1) in the invariant $45 \cdot
\overline{126}\cdot 126$. \bea {1 \over
{2(4!)}}A_{a_{1}a_{2}}\Sigma^{(-)}_{a_{1}i_1..i_4}\Sigma^{(+)}_{a_{2}
i_1..i_4}&=&{A_{\nu}}^{\mu}(\Sigma^{(-)\lambda~\alpha\dot\alpha}_{\mu}
{\Sigma^{(+)}_{\lambda}}^{\nu}_{~\alpha\dot\alpha}-\Sigma^{(-)\nu~\alpha\dot
\alpha}_{\lambda}{\Sigma^{(+)\lambda}_{\mu~~\alpha\dot\alpha}})\nonumber\\
&-&\vec\Sigma^{(-)}_{\mu\nu{R}} \cdot A^{~\nu}_{\sigma} \cdot
\vec\Sigma^{(+) \sigma\mu}_{R}+\vec\Sigma^{(+)}_{\mu\nu{L}} \cdot
A^{~\nu}_{\sigma} \cdot \vec\Sigma^{(-)\sigma\mu}_{L}\nonumber\\
&+
&{A_{\nu}}^{~\mu}\widetilde\Sigma^{(-)\nu\lambda}\Sigma^{(+)}_{\lambda\mu}
\eea

 Note the relative minus sign in the $(15,1,1)_A
(15,2,2)_{\pm}(15,2,2)_{\mp}$ \hfil\break and
$((10,3_{\pm})(\overline{10},3_{\pm})(15,1,1)_A)$ terms due to the
property ${a_{\nu}}^{\mu} \stackrel{D}{\rightarrow}
(-)^{\mu+\nu+1}{a_ {\mu}}^{\nu}$.
The terms containing $A_{\ta\tb}$ are given by %
\bea
{1 \over 4!} A_{\ta\tb}\Sigma_{\ta{i_1..i_4}}^{(-)}\Sigma_
{\tb{i_1..i_4}}^{(+)}
&=&\sqrt{2}\{\vec{A}_{R} \cdot ({\vec\Sigma^{{R}(-)}_{\mu\nu} \times \vec
\Sigma^{(+)\mu\nu}_{R}})+\vec{A}_{L} \cdot ({\vec\Sigma^{\mu\nu(-)}_{L}
\times \vec\Sigma^{(+)L}_{\mu\nu}})\nonumber\\
&-&({A^{\dot\alpha\dot\beta}\Sigma^{(-)\mu\alpha}_{\nu~~~~\dot\alpha}\Sigma^
{(+)\nu}_{\mu~~~\alpha\dot\beta}+A^{\alpha\beta}\Sigma^{(-)\mu~~\dot\beta}
_{\nu~~~\alpha}\Sigma^{(+)\nu}_{\mu~~~\beta\dot\beta}})\}
\eea
The invariance under D parity of both terms follows from the rules 
 (\ref{tendp1},\ref{tendp2}) which imply  
\be
\vec{A}_{R} \cdot ({\vec{B}_{R} \times \vec{C}_{R}}) \leftrightarrow
\vec{A}_{L} \cdot ({\vec{B}_{L} \times \vec{C}_{L}})
\ee
\section {Spinor Representations}
\subsection{Generalities of SO(2N) Spinors}
\noindent In the Wilzcek and Zee \cite{wilzee}  notation  the
$\gamma$ matrices of the Clifford algebra of
SO(2N),~$\gamma{{_i}^{(N)}}$ are defined iteratively as direct
products of Pauli matrices. \bea
\gamma_{i}^{(n+1)}&=&\gamma{{_i}^{(n)}}\otimes \tau_3 ,\quad
n=1.....N-1\\ \gamma_{(2n+1)}^{(n+1)}&=&1\otimes \tau_1\\
\gamma_{(2n+2)}^{(n+1)}&=&1\otimes \tau_2 \eea starting with
$\gamma^{(1)}_{1}=\tau_1 ~,~\gamma^{(1)}_{2}=\tau_2$. One also
defines
\be
\gamma^{(N)}_{F}={(-i)^N}\prod_{i=1}^{2N}\gamma_{i}^{(N)} \equiv
\bigotimes_{i=1} ^{N}{(\tau_3)_i}=\gamma_F^{(m)} \otimes
\gamma_F^{(N-m)} ,~m=1,...N-1 \label{gF} \ee so that
$\gamma_{F}^2=1 ~,~ \gamma_{F}\gamma_{i}=-\gamma_{i}\gamma_{F}$.
\noindent The generators of SO(2N) in the spinor representation
are defined as $(i \not=j)$
\be
J_{ij}=-{\sigma_{ij} \over 2}=-{i \over 4}[\gamma_i,\gamma_j]
\ee
A crucial point (where we disagree with equation (A19) of \cite{wilzee})
 is the form of the charge conjugation matrix C.
Equation A(19) of \cite{wilzee} appears to contradict equation
A(11) of the same paper since ($(-)^n\neq(-)^{n(n+1) \over 2}$ in
general).\\ Recall that $\psi^{T}C\chi$ is a SO(2N) singlet when
\be
\sigma_{ij}^{T}C=-C\sigma_{ij} \ee Two obvious possible (real)
 choices
for C are
\be
C_1^{(n)}=\prod_{j=1}^{n}\gamma_{2j+1}~~~~,\quad
C_2^{(n)}=i^n\prod_{j=1}^{n}\gamma_{2j}
\ee
\be
{\rm{then}}~~~~~~~C_1^{(n)T}={(-)}^{n(n-1) \over 2}C_1^{(n)}~~~~,~~~~
C_2^{(n)T}={(-)}^{n(n+1) \over 2}C_2^{(n)}\label{cT}
\ee
\be
\gamma_{i}^{T}C_{1}=(-)^{n-1}C_{1}\gamma_{i}~~~~~,~~~~
\gamma_{i}^{T}C_{2}=(-)^{n}C_{2}\gamma_{i} \ee and both obey
$C\gamma_F={(-)^n}\gamma_F C$. Their explicit forms are easily
obtained from \bea C_1^{(1)}&=& \tau_1 ~~,~~ C_2^{(1)}=i \tau_2\\
C_1^{(n)}&=& \tau_1\times C_2^{(n-1)}\\ C_2^{(n)}&=& i\tau_2\times
C_1^{(n-1)} \eea \noindent
 In particular $C_2^{(2m+1)}=i\tau_2
\times {\bigotimes}_{i=1}^{m}(\tau_1\times i\tau_2)_{i} $ is
clearly very different from eqn. A(19) of \cite{wilzee} which reads
\be
C=i\tau_2\times i\tau_2 \times i\tau_2 \times \cdot\cdot\cdot
\label{wrongC}\ee
and thus our charge conjugation matrices obey
their eqn. A(11) (our eqn(\ref{cT})) while (\ref{wrongC}) does not.

\noindent On chiral spinor irreps (projected using $({{1\pm
\gamma_{F}}\over 2})$) $C_{1}$ and $C_{2}$ are essentially
equivalent. We shall define the SO(2N) charge conjugation matrix
to be $C_2^{(N)}$. The Clifford algebra of SO(2N) acts on a $2^N$
dimensional space which is given the convenient basis of
eigenvectors ${|\epsilon=\pm 1>}$ of $\tau_3$:
\be
{|\epsilon_1,.......\epsilon_n>}={|\epsilon_1>}\otimes......\otimes{|\epsilon_n>}
\ee
In this basis $\gamma_F=\prod_{i=1}^{n}\epsilon_i$. So the basis spinors
of SO(2N) decompose
into odd and even subspaces w.r.t. $\gamma_F$.
\be
2^n=2_{+}^{n-1}+2_{-}^{n-1}
\ee
The SO(2N) dual of an N index object is
\be
{\widetilde{F}}_{i_1.....i_N}=-{i^N \over
N!}\epsilon_{i_1......i_{2N}}F_{i_{N+1}...i_{2N}}
\ee
The identity
\be
\gamma_{[i_1}......\gamma_{i_M]}\gamma_F={(-i)^{N}(-)^{{M(M-1)}
\over 2}M! \over {(2N-M)!}}
\epsilon_{i_1......i_{2N}}\gamma_{i_{M+1}}.........\gamma_{i_{2N}}\label{gm}
\ee is also frequently needed.
\subsection{SO(6) Spinors}
The $4(\psi_\mu)$ and $\bar 4(\widehat{\psi}^\mu)$ of SU(4) may be consistently
identified with the $4_{-},4_{+}$ chiral spinor multiplets of SO(6) by
identifying components
${\psi_\mu}$ of the 4 with the coefficients of the states
$|\epsilon_1\epsilon_2\epsilon_3>_{-}$
in $4_{-}=|\psi>_{-}$ as
\be
|\psi>_{-}=\psi_{1}|-++> +~\psi_{2}|+-+> +~\psi_3|++-> +~\psi_4|--->
\ee
and also ${\widehat\psi^\mu}$
in the $4_{+}=|\widehat\psi>_{+}$ as
\be
|\widehat\psi>_{+}=-\widehat\psi^{1}|+--> +~\widehat\psi^{2}|-+-> -
~\widehat\psi^{3}|--+> +~\widehat\psi^{4}|+++>
\ee
The reason for
the extra minus signs is that then the charge conjugation matrix
$C_2^{(3)}$ correctly combines the $4,\bar 4$ components in the $2^3$-plet
spinors of SO(6) to make SU(4) singlets and covariants . For example
(we take $\psi,\chi$ to be non-chiral
$8=4_{+}+4_{-}$ spinors to preserve generality)
\be
\psi^{T}\bar C_{2}^{(3)}\chi=\widehat\psi^{\mu}\chi_{\mu}+\psi_{\mu}\widehat
\chi^{\mu}
\ee
\be
\psi^{\dagger}\chi=\psi^{*}_{\mu}\chi_{\mu}+\widehat{\psi}^{\mu*}\widehat\chi^{\mu}
\ee while
\be
D_{abc}^{\pm}\equiv {1 \over
3!}\psi_{\mp}^{T}C_2\gamma_{[a}\gamma_{b}\gamma_{c]}
\chi_{\mp}=\pm\widetilde{D}_{abc}^{\pm}
\ee
\bea
\rm{i.e}~~~~~
{(4_{-} \times 4_{-})_{self-dual}}\leftrightarrow 10~~ of~~ SU(4)\\
{(4_{+} \times 4_{+})_{anti.s.d}}\leftrightarrow {\overline {10}}~~ of
~~SU(4)
\eea
Which is consistent with the identification $4_{-}\sim 4, 4_{+} \sim\bar{4}$
and the multiplication rules in SU(4).
Transforming to the basis in which the components of the spinor
$8=4_{-}+4_{+}$ are precisely the $4+\overline{4}$ i.e.
$(\psi_{\mu},\widehat{\psi}^{\mu})$, one finds that in that basis
\be
C_{2}^{(3)}=AntiDiag(I_{4},I_{4})~,~C_{1}^{(3)}=AntiDiag(I_{4},-I_{4})
\ee

$$ [\gamma_{\mu\nu}]=
\bordermatrix{ & {\{}\}^{\sigma} & {\{}\}_{\sigma}\cr
 {\{}\}_{\lambda} & 0 &
\sqrt{2}\epsilon_{\mu\nu\lambda\sigma} \cr {\{}\}^{\lambda} &
-\sqrt{2}\delta^{\lambda}_{[\mu}\delta^{\sigma}_{\nu]} & 0\cr}
$$
In this basis one has in the 8 dimensional spinor rep. of SO(6) 

$$ exp({{i\omega^{ab}J_{ab}}\over 2}) = Diag {\big (} 
exp({{i\theta^A \lambda^A}\over 2}) , exp({{-i\theta^A \lambda^{A*}}\over 2}){\big )} $$

when the parameters are related as in eqn(\ref{param}).
One finds the following useful identities hold 
\be
\begin{array}{cl}
\psi^{T}{C}_{2}^{(3)}\chi&=\psi_{\mu}{\widehat\chi}^{\mu}+{\widehat\psi}^{\mu}
\chi_{\mu}=\psi.\widehat\chi+\widehat\psi.\chi\\
\psi^{T}{C}_{2}^{(3)}\gamma_{\mu\nu}\chi&=\sqrt{2}{(-\psi_{[\mu}\chi_{\nu]}+
{\widehat\psi}^{\lambda}{\widehat\chi}^{\sigma}\epsilon_{\mu\nu\lambda\sigma})}\\
\psi^{T}{C}_{2}^{(3)}\gamma_{\mu\nu}\gamma_{\lambda\sigma}\chi&=-2{\{\widehat\psi^
{\theta}\chi_{[\lambda}\epsilon_{\sigma]\mu\nu\theta}+\psi_{[\mu}\epsilon_{\nu]
\lambda\sigma\theta}\widehat\chi^{\theta}\}}\\
\psi^{T}{C}^{(3)}_{2}\gamma_{\mu\nu}\gamma_{\lambda\sigma}\gamma_{\theta\delta}
\chi&={(\sqrt{2})^3}{\{\psi_{[\mu}\epsilon_{\nu]\lambda\sigma[\theta}\chi
_{\delta]}+{\widehat\psi}^{\omega}\widehat\chi^{\rho}\epsilon_{\omega\mu\nu
[\theta}\epsilon _{\delta]\rho\lambda\sigma}\}}
\end{array}\label{so6spin}
\ee The results when $\psi^{T}C^{(3)}_{2}\rightarrow
\psi^{\dagger}$ are obtained by the replacements
$\psi_\mu\rightarrow \widehat{\psi}^{\mu*}$  and
$\widehat{\psi}^{\mu}\rightarrow\psi^{*}_{\mu}$ on the R.H.S of
all the identities in (\ref{so6spin}). The square root factors
arise because the antisymmetric pair labels for the gamma matrices
correspond to complex indices ${\hat a}, {\hat b}$. Note that due
to (\ref{gm}) one does not need the identities for more than 3
gamma matrices. See the appendix for useful translations of SO(6)
spinor-tensor invariants calculable from these identities .
\subsection {SO(4) Spinors}
\noindent
In the case of SO(4) the spinor representation is 4 dimensional and
splits into $2_+\oplus 2_-$. It is not hard to see that with the
definitions adopted for the generators of $SU(2)_{\pm}$ the  chiral spinors
$2_{\pm}$ may be identified with the
doublets ${\psi_{\alpha},\psi_{\dot\alpha}}$ of $SU(2)_{-}=SU(2)_L$ and
$SU(2)_{+}=SU(2)_R$ as
\be
|2>_{-}=|\psi>_{-}=\psi_{1}|+-> +~\psi_{2}|-+> ,\quad
|2>_{+}=|\psi>_{+}=\psi_{\dot{1}}|++> -~\psi_{\dot{2}}|--> \ee As
in the SO(6) case one transforms to the unitary basis where
$4=2_{+}\oplus 2_{-}$ has components $(\psi_{\alpha},\psi_{\da})$.
Then in that basis \be C_{2}=\left(\begin{array}{cc}
\epsilon^{\alpha\beta} & 0_{2}\\ 0_{2} & -\epsilon^{\da\db}
\end{array}\right)~,~C_{1}=-\left(\begin{array}{cc}
\epsilon^{\alpha\beta} & 0_{2}\\ 0_{2} &
\epsilon^{\da\db}\end{array}\right)~,
~[\gamma_{\rho\dot\rho}]=\sqrt{2}\left(\begin{array}{cc}
0_{2} & \epsilon_{\rho\alpha}\delta^{\db}_{\dot\rho}
\\\epsilon_{\dot\rho\da}\delta^{\beta}_{\rho} & 0_{2}
\end{array}\right)
\ee

 The following expressions for spinor covariants then follow
\be
\begin{array}{cl}
{\psi}^{T}C_{2}^{(2)}\chi&=\psi^{\dot\alpha}\chi_{\dot\alpha}-\psi^{\alpha}
\chi_{\alpha}\\
{\psi}^{T}C_{1}^{(2)}\chi&=\psi^{\dot\alpha}\chi_{\dot\alpha}+\psi^{\alpha}
\chi_{\alpha}\\
\psi^{T}{C}_{2}^{(2)}\gamma_{\alpha\dot\alpha}\chi&=\sqrt{2}{( \psi_{\dot
\alpha}\chi_{\alpha}-\psi_{\alpha} \chi_{\dot\alpha})}\\
\psi^{T}{C}_{1}^{(2)}\gamma_{\alpha\dot\alpha}\chi&=\sqrt{2}{( \psi_{\dot
\alpha}\chi_{\alpha}+\psi_{\alpha} \chi_{\dot\alpha})}\\
\psi^{T}{C}_{2}^{(2)}\gamma_{\alpha\dot\alpha}\gamma_{\beta\dot\beta}\chi&=
2{\epsilon_{\dot\alpha\dot\beta}\psi_{\alpha}\chi_{\beta}-2\epsilon_{\alpha
\beta} \psi_{\dot\alpha} \chi_{\dot\beta}}\\
%
\psi^{T}{C}_{1}^{(2)}\gamma_{\alpha\dot\alpha}\gamma_{\beta\dot\beta}\chi&=
{-2\epsilon_{\dot\alpha\dot\beta}\psi_{\alpha}\chi_{\beta}-2\epsilon_{\alpha
\beta} \psi_{\dot\alpha} \chi_{\dot\beta}}
\end{array}\label{so4spin}
\ee

Furthermore
\be
\begin{array}{cl}
\psi^{\dagger}\chi&=\psi^{*}_{\da}\chi_{\da}+\psi^{*}_{\alpha}\chi_{\alpha}\\
\psi^{\dagger}\gamma_{\alpha\da}\chi&=-{\sqrt{2}}(\psi^{\alpha*}\chi_{\da}+\psi^{\da*}\chi_{\alpha})\\
\psi^{\dagger}\gamma_{\alpha\da}\gamma_{\beta\db}\chi&=2\epsilon_{\da\db}
\psi^{\alpha*}\chi_{\beta}+2\epsilon_{\alpha\beta}\psi^{\da*}\chi_{\db}
\label{so4spin2} 
\end{array}
\ee Note that these can be obtained from the corresponding
identities involving $C^{(2)}_{1}$ by the replacements
$\psi^{\da}\rightarrow \psi^{*}_{\da},\psi^{\alpha}\rightarrow
\psi^{*}_{\alpha}$ or from the $C_{2}$ identities by
$\psi^{\da}\rightarrow
\psi_{\da}^{*},\psi^{\alpha}\rightarrow-\psi_{\alpha}^{*}$.
\subsection {SO(10) Spinors}
The spinor representation of SO(10) is $2^5$ dimensional and
 splits into chiral eigenstates with $\gamma_F=\pm 1$ as
\bea
2^5&=&2^{4}_{+}+2^{4}_{-}=16_{+}+16_{-}\\
16&=&16_{+}=(4_{+},2_{+})+(4_{-},2_{-})=(\overline{4},1,2)+(4,2,1) \\
\overline{16}&=&16_{-}=(4_{+},2_{-})+(4_{-},2_{+})=(\overline{4},2,1)+
(4,1,2)
\eea
Where the first equality follows from eqn(\ref{gF}) and second from the SO(6)
to SU(4) and SO(4) to $SU(2)_{L} \times SU(2)_{R}$ translations: $4_-=4,2_+=2_{R}
,2_-=2_{L}$.
Thus we see that the SU(4) and $SU(2)_L \times SU(2)_R$ properties of the
submultiplets  within the $16,\overline{16}$ are strictly correlated.
Use of the SO(6) and SO(4) spinor covariant identities
allows fast construction of SO(10) spinor invariants.
For example ,
\be
\psi^{T}C_{2}^{(5)} \gamma_{\mu\nu}^{(5)}\chi = \psi^{T}(C_2^{(3)}
\times C_{1}^{(2)})(\gamma_{\mu\nu}^{(3)} \times \tau_3 \times
\tau_3)\chi =\psi^T(C_2^{(3)}\gamma_{\mu\nu}^{(3)}\times
C_2^{(2)})\chi \ee Next one uses the identities
(\ref{so6spin},\ref{so4spin}) in parallel , keeping in mind that
in the 16-plet the dotted ($\rm{SU(2)}_R $) spinors are always $\bar
4$-plets of SU(4) and the undotted ones are 4-plets and vice versa
for $\overline{16}$ . When $\psi,\chi$ are both 16-plets one
immediately reads off the result
\be
\psi^{T}C_{2}^{(5)}\gamma_{\mu\nu}^{(5)}\chi=\sqrt {2}{(\psi^{\alpha}_{[\mu}\chi_{{\nu]}\alpha}+
{\widehat\psi}^{\lambda{\dot\alpha}}{\widehat\chi}^{\sigma}_{\dot\alpha}
\epsilon_{\mu\nu\lambda\sigma})}
\ee

{\bf{\underline{D parity on spinors}}}~:~
\vspace{.1 cm}\noindent
D parity acts on the spinors of SO(10) as
\bea
D_{spinor}&=&e^{({-i{\pi}J_{23}})}e^{({i{\pi}J_{67}})}=-\gamma_2\gamma_3\gamma_6
\gamma_7\nonumber\\
&=&(\bigotimes_{i=1}^{3}i\tau_2) \times (i\tau_2 \times 1_2)=D^{(3)} \times
 D^{(2)}
\eea
Thus the action of D factorizes. Under $D^{(3)}$
one interchanges spinors of opposite chirality as :
\be
\widehat\psi^\mu \rightarrow (-)^{\mu + 1}\psi_\mu
\ee
\be
\psi_\mu \rightarrow (-)^{\mu}\widehat\psi^{\mu}
\ee
Similarly for $D^{(2)}=i\tau_2 \times 1$, one finds interchange
\be
{\psi}_{\alpha} \rightarrow {\psi}_{\dot{\bar\alpha}}  ,
\quad {\psi}_{\dot\alpha}
\rightarrow {-\psi}_{\bar\alpha} 
\Rightarrow \psi^{\alpha} \rightarrow -\psi
^{\dot{\bar\alpha}},\psi^{\dot\alpha} \rightarrow +\psi^{\bar\alpha}
\ee
Where by $\bar\alpha$ we mean $\bar{1}=2,\bar{2}=1.$
This implies the contraction of spinors $\psi_{\alpha},\chi_{\dot\alpha}$ with
a bidoublet $V_{\alpha\dot\alpha}=V_{\hat{a}}$ tranforms  as
\be
V^{\alpha\dot\beta}\psi_{\alpha}\chi_{\dot\beta} \rightarrow -V^{\bar\beta
\dot{\bar\alpha}}\psi_{\dot{\bar\alpha}}\chi_{\bar\beta}
\ee
Similarly with $\rm{SU(2)}_{L} (\rm{SU(2)}_{R})$ vectors one gets
\bea
V_{(-)}^{\alpha\beta}\psi_{\alpha}\chi_{\beta} &\leftrightarrow&
-V_{(+)}
^{\dot{\bar\alpha}\dot{\bar\beta}}\psi_{\dot{\bar\alpha}}\chi_{\dot{\bar
\beta}} \\ {\rm{While}}~~~~~~~~~~~~~~~~~~~~~ \psi^{\alpha}\chi_{\alpha} &\leftrightarrow&
-\psi^{\dot\alpha}\chi_{\dot \alpha}\\
\widehat\psi^{\mu}\chi_{\mu} &\leftrightarrow&
-\psi_{\mu}\widehat\chi^{\mu} \eea

These rules are consistent with the action of D-parity 
on PS subreps SO(10) tensors derived earlier . Indeed one 
recovers them when one defines such  tensors via bilinear
covariants formed from SO(10) spinors.

{\bf{\underline{SO(10) Spinor-Tensor Invariants  }}}
\vspace{.1 cm}\\
We next give the explicit decomposition of  quadratic and cubic
SO(10) invariants involving a pair ($16,16$ or $16,\overline {16}$ ) of SO(10)
spinors contracted with (the conjugate of) one of the tensors in their
Kronecker product decomposition :
\bea
16\otimes 16 &=& 10 \oplus 120\oplus 126\\
16\otimes {\overline{16}} &=& 1 \oplus 45 \oplus 210
\eea
Besides use of the spinor identities (\ref{so6spin},\ref{so4spin})
the remainder of the task is merely to decompose the SO(10) index
contractions into PS irrep. index contractions , take account of
self-duality where relevant and maintain unit reference norm.
\vspace{.1cm}\\
${\bf{\underline{16 \cdot 16 \cdot 10}}}$~:~
\vspace{.1 cm}
The 10-plet has decomposition:
$H_{i}(10)=H_{a}{(6,1,1)}+H_{\ta}{(1,2,2)}$ and one gets
\be
\psi^{T}{C}_{2}^{(5)}\gamma_{i}^{(5)}\chi H_{i}=\sqrt{2}
{\{H_{\mu\nu}\widehat{\psi}^{\mu{\dot
\alpha}}\widehat{\chi}^{\nu}_{\dot\alpha}+
\widetilde{H}^{\mu\nu}\psi_{\mu}^{\alpha}\chi_{\nu\alpha}-H^{\alpha\dot\alpha}
{(\widehat{\psi}^{\mu}
_{\dot\alpha}\chi_{\alpha\mu}+\psi_{\alpha\mu}\widehat{\chi}_{\dot\alpha}^{\mu})}}\}\\
\label{psi10}\ee
Note how D parity is maintained by the interplay between the SO(6) and SO(4)
sectors.
\vspace{.1cm}\\
${\bf{\underline{16 \cdot 16 \cdot 120}}}$~:~
\vspace{.1 cm}
Since
\bea
O_{ijk}(120)&=&O_{abc}(10+{\overline{10}},1,1)+O_{ab \ta}(15,2,2)+O_{a\ta\tb}((6,1,3)+
(6,3,1))+ O_{\ta\tb\tilde\gamma}(1,2,2)\nonumber\\
&=&O^{(s)}_{\mu\nu}({10,1,1})+\overline{O}^{\mu\nu}_{(s)}(\overline{10},1,1)
+{O_{\nu\alpha\dot\alpha}}^{\mu}(15,2,2)\nonumber\\
&+& O^{(a)}_{\mu\nu\dot\alpha\dot\beta}(6,1,3)+
{O^{(a)}_{\mu\nu}}_{\alpha\beta}(6,3,1)+
O_{\alpha\dot\alpha}(1,2,2) \eea (where we have used the
superscripts ${}^{(s),(a)}$ to discriminate the symmetric 10-plet
from the antisymmetric 6-plet). Then one gets
 \bea {1 \over
(3!)}\psi C^{(5)}_{2}\gamma_{i}\gamma_{j}\gamma_{k}\chi O_{ijk}&
=& -2{(\bar{O}^{\mu\nu}_{(s)}\psi_{\mu}^{\alpha}\chi_{\nu\alpha}
+O_{\mu\nu}^{(s)}\widehat\psi^{\mu\dot\alpha}{\widehat\chi_{\dot\alpha}^{\nu}}
)}\nonumber\\
&-&2\sqrt{2}{O^{~\mu\alpha\dot\alpha}_{\nu}}
{{(\widehat\psi_{\dot\alpha}^{\nu}\chi_{\mu\alpha}-
{\psi_{\mu\alpha}\widehat\chi_{\dot\alpha}^\nu})}} \nonumber\\
&-&2({O_{\mu\nu}^{(a)}}^{\dot\alpha\dot\beta}
\widehat\psi^{\mu}_{\dot\alpha}\widehat\chi^{\nu}_{\dot\beta}+\widetilde{O}^
{\mu\nu\alpha\beta}_{(a)}\psi_{\mu\alpha}\chi_{\nu\beta})\nonumber\\
&+&\sqrt{2}O^{\alpha
\dot\alpha}(+\hat\psi_{\dot\alpha}^{\mu}\chi_{\mu\alpha}-\psi_{\mu\alpha}
\widehat\chi_{\dot\alpha}^{\mu})
\eea
Note $O^{\alpha\dot\alpha}$ is derived from $O_{\tilde\alpha}=-{1
\over 3!}
\epsilon_{\ta\tb\tilde\gamma\tilde\delta}O_{\tb\tilde\gamma\tilde\delta}$
and so has opposite D parity to a vector $V_{\ta}$.
\vspace{.1cm}\\
${\bf{\underline{16 \cdot 16 \cdot \overline{126}}}}$
\be
\overline{126}={\overline\Sigma}_{\mu\nu}^{(a)}(6,1,1)+{{\overline\Sigma}_{\nu}^{~\mu}}_{\alpha\dot\alpha}
(15,2,2)+{\overline\Sigma}_{\mu\nu,\dot\alpha\dot\beta}(10,1,3)+{\overline{{\overline\Sigma}}^{\mu\nu}}_{\alpha\beta}
({\overline{10}},3,1) \ee
  \bea {1 \over
5!}\psi^{T}C_{2}^{(5)}\gamma_{i_{1}}.....\gamma_{i_{5}}\chi
{\overline\Sigma}_{i_{1}...{i_{5}}}&=&2\sqrt
{2}{(\widetilde{{\overline\Sigma}}^{\mu\nu}_{(a)}\psi_{\mu}^
{\alpha}\chi_{\nu\alpha}-{\overline\Sigma}_{\mu\nu}^{(a)}\widehat
\psi^{\mu\dot\alpha}{\widehat
 \chi}_{\dot\alpha}^{\nu})}\nonumber\\
&+&{4 \sqrt {2}}{\overline\Sigma}^{~\mu\alpha\dot\alpha}_{\nu}
 {(\widehat\psi_{\dot\alpha}^{\nu}\chi_{\alpha\mu}+\psi_{\mu\alpha}\widehat
\chi_{\dot\alpha}^{\nu})}\nonumber\\
&+&4({\overline\Sigma}_{\mu\nu}^{\dot\alpha\dot\beta}\widehat\psi_{\dot\alpha}^{\mu}
\widehat\chi_{\dot\beta}^{\nu}+{{\overline\Sigma}}^{\mu\nu,\alpha\beta}\psi_{\mu\alpha}
\chi_{\nu\beta})\label{psi126} \eea
Here $({\overline\Sigma}_{\mu\nu}^{(a)}) \leftrightarrow
(-)^{\mu+\nu}\widetilde{{\overline\Sigma}}^{\mu\nu} _{(a)} ,~~ {{\overline\Sigma}_{\mu}}^{\nu}
\leftrightarrow (-)^{\mu+\nu}{{\overline\Sigma}_{\nu}}^{\mu}$ have reversed D
parity due to the dualization involved in their definition. We say
a representation is D-Axial if due to dualization it has an extra
minus sign in its D transformation relative to that expected from
its tensor structure.\\ \vspace{.1cm}
\vfil\eject
${\bf{\underline{16 \cdot \overline{16} }}}$
\bea
16(\psi)=(4,2,1)\psi_{\mu\alpha}+(\bar{4},1,2)\hat\psi^{\mu}_{\dot\alpha}\\
{\overline{16}} (\overline\chi)=(\bar{4},2,1)\hat\chi^{\mu}_{\alpha}+(4,1,2)\chi_{\mu\dot\alpha}
\eea
\be
\psi^T{C}_{2}^{(5)}\overline\chi=\widehat\psi^{\mu\dot\alpha}\chi_{\mu\dot
\alpha}+\psi_{\mu\alpha}\widehat\chi^{\mu\alpha}=-\chi^{T}C_{2}^{(5)}\psi
\ee
${\bf{\underline{16 \cdot \overline{16} \cdot 45}}}$
\be
45={A_{\nu}}^{\mu}(15,1,1)+A_{\mu\nu,\alpha\dot\alpha}(6,2,2)+A_{\alpha
\beta}(1,3,1)+A_{\dot\alpha\dot\beta}(1,1,3)\ee\bea   {1 \over
(2!)}\psi^{T}{C}_{2}^{(5)}\gamma_{i}\gamma_{j}\overline{\chi}
{A_{ij}}
&=&2A_{\kappa}^{~\mu}(-\psi_{\mu}^{\alpha}\widehat\chi_{\alpha}^{\kappa}+
\widehat\psi^{\kappa\dot\alpha}\chi_{\mu\dot\alpha})\nonumber\\
&-&\sqrt2{(A^{\dot\alpha
\dot\beta}\widehat\psi^{\mu\dot\alpha}\chi_{\dot\beta\mu}+A^{\alpha\beta}
\psi_{\mu\alpha}\widehat\chi^{\mu}_{\beta}})\nonumber \\
&-& {(\widetilde{A}^{\mu\nu,\alpha\dot\alpha}\psi_{\mu\alpha}\chi_{\nu\dot
\alpha}+A_{\mu\nu}^{~\alpha\dot\alpha}\widehat\psi^{\mu}_{\dot
\alpha}\widehat\chi_{\alpha}^{\nu})}
\eea
${\bf{\underline{16 \cdot \overline{16} \cdot 210}}}$:
\bea
210&=&{{\Phi}_{\nu}}^{\delta}(15,1,1)+{\Phi}_{\mu\nu,\alpha\dot\alpha}(10,2,2) +
{\overline{{\Phi}}^{\mu\nu}}_{\alpha\dot\alpha}(\overline{10},2,2)\nonumber\\
&+&{\Phi}^{~\nu}_{\mu,\alpha\beta}(15,3,1)+
{\Phi}^{~\nu}_{\theta,\dot\alpha\dot\beta}(15,1,3)+{\Phi}(1,1,1) \eea
\bea {1 \over
(4!)}\psi^{T}{C}_{2}^{(5)}\gamma_{i_{1}}....\gamma_{i_{4}}
\overline{\chi}{{\Phi}}_ {i_{1}...i_{4}}
&=&-2i{{\Phi}}^{~\sigma}_{\delta}(\widehat\psi^{\delta\dot\alpha}\chi_
{\sigma\dot\alpha}+\psi_{\sigma\alpha}\widehat\chi^{\delta}_
{\alpha}) \nonumber\\
&+&2
\sqrt{2}(\overline{{\Phi}}^{\mu\nu,\alpha\dot\alpha}\psi_{\mu\alpha}\chi_
{\nu\dot\alpha}+{{\Phi}_{\mu\nu}}^{\alpha\dot\alpha}\widehat\psi^{\mu}_{\dot\alpha}\widehat\chi^
{\nu}_{\alpha}) \nonumber\\
&+&2
\sqrt{2}{\{{{\Phi}_{\delta}}^{\mu,}}^{\alpha\beta}\psi_{\mu\alpha}{\widehat
\chi^{\delta}}_{\beta}-{{{\Phi}_{\delta}}^{\lambda\dot\alpha\dot\beta}
\widehat\psi^{\delta}_{\dot\alpha}\chi_{\lambda\dot\beta}\}}\nonumber\\
&+&2{\{\widetilde{{\Phi}}^{\mu\nu,\alpha\dot\alpha}\psi_{\mu\alpha}\chi_{\nu\dot
\alpha}+{{\Phi}_{\mu\nu}}^{\alpha\dot\alpha}
\widehat\psi^{\mu}_{\dot\alpha}{\widehat\chi^{\nu}}_{\alpha}\}}\nonumber\\
& &+{\Phi}(\psi_{\mu}^{\alpha}{\widehat\chi^{\mu}}_{\alpha}
-\widehat\psi^{\mu\dot\alpha}\chi_{\mu\dot\alpha})
\eea
${\Phi}^{~\mu}_{\nu} , {\Phi}$  are both D-Axial, while
\be
D({\Phi}_{\mu\nu}^{~\alpha\dot\beta})=(-)^{\mu+\nu+1}\overline{{\Phi}}^{\mu\nu\bar\beta
\dot{\bar\alpha}}
\ee
Note that to obtain the results when $16^*$ is used instead of $\overline{16}$
one need only replace
\be
\widehat\chi^{\mu\alpha} \rightarrow \chi^*_{\mu\alpha}  ,\quad
{\chi_{\mu}}^{\dot\alpha} \rightarrow
({\widehat\chi^{\mu}}_{\dot\alpha})^* \ee because
$C_{2}^{(5)}=C_{2}^{(3)}\times C_{1}^{(2)}$ (see the remarks following 
eqns(\ref{so6spin},\ref{so4spin2}). When calculating
quartic invariants formed by contractions of SO(10) tensor
covariants
made from 16, ${\overline{16}}$ multiplets (which often arise in model building
with non renormalizable superpotentials \cite{so10refs}) one need only
apply the identities (\ref{so6spin},\ref{so4spin}) after decomposing the SO(10) vector
 indices while treating one of the covariants as an operator with appropriate
PS indices.
\section{  Illustrative Applications }
In this section we give some examples of the use of our methods for typical tasks that arise 
when studying GUTs. The first illustration is a the translation of the SO(10) 
covariant derivative to PS labels. The second is an an explicit calculation 
of the bare effective potential for Baryon Decay in the ``minimal Susy GUT "
\cite{abmsv} which is of direct phenomenological interest  and constitutes the main physical 
result of this paper.

The translation of the SO(10)  covariant derivatives may be seen from e.g.
\bea
{\psi}^\dagger({\partial}+{i \over 2}{A}^{kl}g_{u}J_{kl})\psi&=&
{\psi}_{\mu\alpha}^*\partial\psi_{\mu\alpha}+
{\widehat\psi^{\mu *}
_{\dot\alpha}}\partial\widehat\psi_{~\dot\alpha}^{\mu}\nonumber\\
& &+{ig_u \sqrt2}\{\psi_{\kappa\alpha}^*{A}^{A}({\lambda^{A}
\over 2})_{\kappa\mu}\psi_{\mu\alpha}+
{\widehat\psi^{\mu *}}_{~~\dot\alpha}{A}^{A}
({({-\lambda_{A} \over 2})_{\mu\kappa}})^*\widehat\psi^{\kappa}_{~~\dot\alpha}
\nonumber\\
& &+{{\widehat\psi}^{\mu *}_{\dot\beta}}({\vec{ A}_{R} \cdot \vec\sigma \over 2})
_{\dot\beta}^{~~\dot\gamma}\widehat\psi_{\mu\dot\gamma}
+{\psi_{\mu\beta}}^*({\vec{A}_{L} \cdot \vec\sigma \over 2})
_{\beta}^{~~\gamma}\widehat\psi_{\mu\gamma}\}\nonumber\\
& &+{g_{u} \over 2}({\widehat\psi^{\nu *}_{\dot\alpha}}\widetilde{A}
^{\mu\nu\alpha}_{~~\dot\alpha}\psi_{\mu\alpha}+\psi_{\nu\alpha}^*
{A}_{\mu\nu\alpha}^{~~\dot\alpha}\widehat\psi^{\mu}_{~~\dot\alpha})
\label{covder}
\eea

We see that Pati-Salam coupling constants emerge as $g_{4}=g_{2}={g_u
\sqrt2}$. The GUT generators $T^{A},\vec{T}_{R},\vec{T}_{L}$
are each normalized to 2 on the 16-plet and have $\sqrt2{g_u}$ as their associated
coupling.  In the vector representation covariant derivative behaves as
\bea
{V}_{i}^*(\partial+{i \over 2}g_u{A}^{kl}J_{kl})_{ij}V_{j}&=&
{1 \over 2}{V}_{\mu\nu}^*\partial{V}_{\mu\nu}+{i \over 2}{g_u
\sqrt2}{V}_{\mu\nu}^*{A}^{A}({\lambda^A \over 2})_{[\mu}^{~~\sigma}V_
{\nu]~\sigma}\nonumber\\
&+&ig_u\sqrt{2}{\overline{V}_{\alpha\dot\alpha}}({\vec{W}_{L}\cdot({\vec\sigma
 \over 2})_{\alpha}^{~~\beta}V_{\beta\dot\alpha}+\vec{W}_{R}\cdot({\vec\sigma \over
2})_{\dot\alpha}^{~~\dot\beta}V_{\alpha\dot\beta}})~~~~~~~~~ \eea
This can easily be adapted to decompose the kinetic terms of any
of the tensor representations.

\subsection{ Baryon Decay }

We further illustrate the application and utility of our methods
by calculating two important mass matrices in the  
 minimal Supersymmetric SO(10) GUT  (\cite{aulmoh},\cite{ckn}
\cite{heme,lee,abmsv}).  A part, but not all, of these matrices
was available earlier using  the results of  \cite{lee} on CG coefficients involving singlet 
subreps of SO(10). However our methods also allow calculations of CG coefficients 
that are not of the restricted class studied in \cite{heme,lee}.
\noindent The chiral supermultiplets of the model consist of a 210-
plet $\Phi_{ijkl}$ responsible for breaking SO(10) down to
$\rm{G}_{3211}= \rm{SU(3)}_{C}\times \rm{SU(2)}_{L}\times \rm{U(1)}_{R}\times
\rm{U(1)}_{B-L}$. A $\overline{126}(\overline{\Sigma}), 126(\Sigma)$
pair is required to be present together to break $U(1)_{R}\times
U(1)_{B-L}\rightarrow U(1)_{Y}$ while preserving Susy and is
capable of generating realistic neutrino masses and mixings via
the type I or type II seesaw mechanisms {\cite{seesaw,mohsen}}.
Moreover the SU(2) doublets in the $\overline{126}+126$ can also
participate in the electroweak symmetry breaking. Finally there is
a 10-plet containing SU(2)$_{L}$ doublets and SU(3) triplets and 3
families of matter contained in 16-plets. The complete
superpotential of this model is given by :

 \bea W&=&{m \over
{2(4!)}} \Phi_{ijkl}\Phi_{ijkl}+{\lambda \over
{{4!}}}\Phi_{ijkl}\Phi_{klmn} \Phi_{mnij}+{M \over
{2(5!)}}\Sigma_{ijklm}\overline\Sigma_{ijklm}\nonumber\\ &+&{\eta \over
4!}\Phi_{ijkl}\Sigma_{ijmno}\overline\Sigma_{klmno}+
{1\over{4!}}H_{i}\Phi_{jklm}(\gamma\Sigma_{ijklm}+
\overline{\gamma}\overline{\Sigma}_{ijklm})\nonumber\\
&+& {M_{H} \over{2}}
H^{2}_{i}+h_{AB}'\psi^{T}_{A}C^{(5)}_{2}\Gamma_{i}\psi_{B}H_{i}+{f_{AB}'\over
5!}\psi^{T}_{A}C^{(5)}_{2}{\gamma_{i_{1}}}...
{\gamma_{i_5}}\psi_{B}\overline{\Sigma}_{i_1...i_5} \eea
The GUT scale vevs that break the gauge symmetry down to the SM
symmetry are {\cite{aulmoh,ckn}}:
\begin{itemize}
\item{i)}
\be
{\langle(15,1,1)\rangle}_{210}:\langle{\phi_{abcd}}\rangle={a\over{2}}
\epsilon_{abcdef}\epsilon_{ef} \ee
   where~
$[\epsilon_{ef}]=Diag(\epsilon_{2},\epsilon_{2},\epsilon_{2}),~~\epsilon_{2}=i\tau_{2}
$. Defining
\be
\phi_{ab}\equiv{1\over{4!}}\epsilon_{abcdef}\phi_{cdef}\ee
 We have
in SU(4) notation $[\phi_{\nu}^{~\lambda}]$ for the (15,1,1) and
\be
[\langle{\phi_{\nu}^{~\lambda}}\rangle]={ia
\over{2}}Diag(I_{3},-3)\equiv {ia\Lambda \over {2}} \ee
 \item{ii)}
\be
\langle(15,1,3)\rangle_{210}~:~\langle\phi_{ab\ta\tb}\rangle=\omega
\epsilon_{ab}\epsilon_{\ta\tb}
\ee
 which translates to
\be
\langle(\vec{\phi}^{(R)\nu}_{\mu~\dot{1}\dot{2}})\rangle=-{\omega\Lambda\over
\sqrt{2}}\equiv {i\langle(\vec{\phi}^{(R)\nu}_{\mu})_{0}\rangle}
\ee
 \item{iii)}
\be
\langle(1,1,1)\rangle_{210}~:~\langle\phi_{\alpha\beta\gamma\delta}\rangle=p\epsilon_{\alpha\beta\gamma\delta}\ee
\item{iv)}
 \be
 \langle(10,1,3)\rangle_{\overline{126}}~:~\langle\overline{\Sigma}_{\hat{1}\hat{3}\hat{5}
\hat{8}\hat{0}}\rangle=
\bar\sigma=-i\langle\overline{\Sigma}^{(R)}_{44(+)}\rangle={\overline{\Sigma}_{44\dot{1}\dot{1}}
\over{\sqrt{2}}} \ee
 \item{v)}
\be
\langle(\overline{10},1,3)\rangle_{{126}}~:~\langle{\Sigma}_{\hat{2}\hat{4}\hat{6}\hat{7}\hat{9}}\rangle=
\sigma=i
\langle{\Sigma}^{(R)44}_{(-)}\rangle={\Sigma^{44}_{\dot{2}\dot{2}}
\over{\sqrt{2}}}\ee \end{itemize} Under the SM gauge group
$G_{231}$ the 10 plet decomposes as
\be
10=H_{\alpha}(2,1,1)+\overline{H}_{\alpha}(2,1,-1)+t^{(1)}_{\bar{\mu}}(1,3,{-2\over3})
+\bar{t}^{\bar\mu}_{(1)}(1,\bar{3},{2\over3}) \ee which are the
doublets and triplets familiar from SU(5) unification. In the case
of SO(10) there are many other types of $G_{321}$
multiplets beyond the ones encountered in the SU(5) case but we
focus here only on the multiplets that can mix with the components
of the 10-plet i.e. those that transform as
H,$\overline{\rm{H}}$,t or $\overline{\rm{t}}$.
The doublet $(2,1,\pm{1})$ sector in fact consists of 4 pairs of
 doublets which are
 \be
h^{(1)}_{\alpha}=H_{\alpha\dot{1}}~,~h^{(2)}_{\alpha}=\overline{\Sigma}_{\alpha\dot{1}}~,~
{h}^{(3)}_{\alpha}={\Sigma}_{\alpha\dot{1}}~,~{h}^{(4)}_{\alpha}
={{\Phi^{44\dot{1}}_{\alpha}}\over {\sqrt{2}}}
\ee where
${\Sigma}_{\alpha\dot\alpha},\overline{\Sigma}_{\alpha\dot\alpha}$
refer to the B-L singlet inside the (15,2,2) submultiplets of the
126,$\overline{126}$ and $h^{(4)}$ comes from the
$(\overline{10},2,2)\subset 210$. Similarly one has \be
\bar{h}^{(1)}_{\alpha}=H_{\alpha\dot{2}}~,~\bar{h}^{(2)}_{\alpha}=\overline{\Sigma}
_{\alpha\dot{2}}~,~
 \bar{h}^{(3)}_{\alpha}={\Sigma}_{\alpha\dot{2}}~,
~\bar{h}^{(4)}_{\alpha}={{\Phi_{44\alpha}^ {~\dot{2}}}\over{\sqrt{2}}}
 \ee On the other hand, there are 5 pairs of triplets t$(1,3,-{2\over{3}}),\overline{\rm{t}}(1,\bar{3},{2\over{3}})$
that mix :
\be
t^{(1)}_{\bar\mu}=H_{\bar\mu4}~,~t^{(2)}_{\bar\mu}={\overline{\Sigma}}^{(a)}_{\bar\mu4}~,~t^{(3)}_
{\bar\mu}=
\Sigma^{(a)}_{\bar\mu4}~,~t^{(4)}_{\bar\mu}=(\vec{\overline{\Sigma}}^{(R)}_{4\bar\mu})_{0}~,~
t^{(5)}_{\bar\mu}=(\vec{\Phi}^{(R)4}_{\bar\mu})_{(-)}\ee
\be
\bar{t}_{(1)}^{\bar\mu}=\widetilde{H}^{\bar\mu4}~,~\bar{t}_{(2)}^{\bar\mu}=\overline
{\Sigma}_{(a)}^ {\bar\mu4}~,~{\bar t}_{(3)}^ {\bar\mu}=
{\Sigma}_{(a)}^{\bar\mu4}~,~\bar{t}_{(4)}^{\bar\mu}=(\vec{\Sigma}_{(R)}^
{4\bar\mu})_{0}~,~
\bar{t}_{(5)}^{\bar\mu}=(\vec{\Phi}_{4(R+)}^{~\bar\mu})\ee
Here $t^{(2)(3)},\overline{t}^{(2)(3)}$ come from the (6,1,1)
content of the $\overline{126}$ and 126 while $t^{(4)},
\overline{t}^{(4)}$ come from $(10,1,3)_{\overline{126}}~
\rm{and}~ (\overline{10},1,3)_{126}$. Finally $t_{(5)}~ \rm{and}~
\overline{t}_{(5)}$ come from $(15,1,3)_{210}$.

\noindent The GUT scale vevs described above give rise to
mass matrices dependent only on the 7
parameters m,M,$M_H,\lambda,\eta,\gamma,\overline\gamma$. A fine
tuning is then required to keep one pair of doublets light while
all the other Higgs are superheavy. The feasibility of this fine
tuning and the determination of the mixtures that stay light
requires explicit calculation of these mass matrices. Our method
allows straightforward and unambiguous calculation of these mass
matrices (as well as all other submultiplet Clebsches ).

\noindent The $h,\overline{h}$ mass matrix can be read off from
the bilinear terms in the superpotential which have the structure
$m_{ij}\overline{h}^{(i)\alpha}h^{(j)}_{\alpha}$. For example the
14 element involves $H_{\alpha\dot{2}}\subset H_{\tilde\alpha}
\rm{and}~ \Phi^{44~\dot{1}}_{\alpha}\subset
(\overline{10},2,2)\sim \phi^{(-)}_{abc\ta}$ and can receive a
contribution only from the term
$\overline{\gamma}\langle\overline{\sigma}\rangle\Phi H$ in W
.i.e. from \bea
-{4\overline{\gamma}\over{4!}}H_{\ta}\Phi_{abc\tb}\langle\overline{\Sigma}_{abc\ta\tb}\rangle
&=&
-{1\over{12}}H_{\ta}\Phi^{(-)}_{abc\tb}\langle\overline{\Sigma}_{abc\ta\tb}^{(+,+)}\rangle
=-{\overline{\gamma}\over{2}}\Phi^{\mu\nu\dot\beta}_{\alpha}
\langle\overline{\Sigma}_{\mu\nu\dot\alpha\dot\beta}
^{(+,+)}\rangle{H}^{\alpha\dot\alpha}\nonumber\\&=&-{\sqrt{2}\over{2}}\overline{\gamma}H^{\alpha\dot{1}}
\Phi^{44\dot{1}}
\overline{\sigma}=-{\overline{\gamma}\over{\sqrt{2}}}\overline{\sigma}
\overline{h}^{\alpha}_{(1)}h_{(4)\alpha} \eea

 \hfil\break
\hfil\break
In this way, by a routine use of the translation identities given in the text and
 in the appendix, one obtains the required "Clebsch-Gordon" coefficients without any ambiguity.

\bea {\cal{D}}=\left({\begin{array}{cccc} -M_{H} &
+\overline{\gamma}\sqrt{3}(\omega-a) & -{\gamma}\sqrt{3}(\omega+a)
& -{\bar{\gamma}{\bar{\sigma}} }\\
-\overline{\gamma}\sqrt{3}(\omega+a) & 0 & -(M+4\eta(a+\omega)) &
0\\ \gamma\sqrt{3}(\omega-a) & -(M+4\eta(a-\omega)) & 0 &
-{2\eta\overline{\sigma}\sqrt{3}}\\ -{\sigma\gamma } &
-{2\eta\sigma\sqrt{3}} & 0 & {-{m}+6\lambda(\omega-a)}
\end{array}}\right) \eea

 The element 43 and 24 are zero since they involve SU(4)
contributions
$\Phi^{(+)}\Sigma\langle\overline{\Sigma}^{(++)}\rangle$ and
$\Phi^{(-)}\overline{\Sigma}\langle\Sigma^{(-+)}\rangle$ between
two 10-plets or two $\overline{10}$-plets which vanish.\\

\noindent In a similar way one can calculate the triplet mass
matrix

\bea {\cal{T}}= \left({\begin{array}{ccccc} M_{H} &
\overline{\gamma}(a+p) & {\gamma}(p-a) & {2\sqrt{2}i
\omega{\bar\gamma}} & i\bar{\sigma}\bar{\gamma}\\ \bar\gamma(p-a) &
0 & M & 0 & 0\\ \gamma(p+a) & M & 0 & 4\sqrt{2}i \omega\eta &
2i{\eta\overline{\sigma}}\\ -2\sqrt{2}i \omega\gamma  & -4\sqrt{2}i
\omega\eta & 0 & M+2\eta{p}+2\eta a &
-2\sqrt{2}{\eta\overline{\sigma}}\\ i\sigma\gamma & 2i \eta\sigma
& 0 & 2\sqrt{2}\eta\sigma & -m - 2\lambda(a+p-4 \omega)
\end{array}}\right)
\eea

These mass matrices are crucial to the phenomenological implications of this
model. The fine tuning condition required to retain one pair of light doublets 
in the effective theory is simply $ det {\cal{D}} =0$ . 
The couplings of these light doublets to matter are then specified
in terms of the  $h^{(1)},h^{(2)},{\bar{h}}^{(1)},{\bar{h}}^{(2)}$
content of the light eigenstates of the doublet mass matrices since only the doublets 
coming from the ${\bf{10}}$ and ${\bf{\overline{126}}}$ couple to light matter fermions
contained in the ${\bf{16}}  $. Furthermore the bare effective superpotential relevant to 
baryon decay can be easily calculated  in terms of 
${\cal S}= {\cal T}^{-1} $ 
by using eqns.(\ref{psi10}),(\ref{psi126})  and the standard PS embedding 

\bea  (4,2,1) = (Q_{\alpha},L_{\alpha} ) \qquad \qquad ({\overline{4}},1,2) =
 ({\overline{Q}}_{\alpha},{\overline{L}}_{\alpha}) \eea

with 
\begin{equation} 
Q=\left({\begin{array}{c}U\\D\end{array}}\right)
 \quad L=\left({\begin{array}{c} \nu\\e\end{array}}\right) \quad
{\overline Q}=\left({\begin{array}{c} {\bar d}\\{\bar u}\end{array}}\right) \quad 
{\overline{L}}=\left({\begin{array}{c}{\bar e}\\{\bar \nu}\end{array}}\right) \quad
\end{equation}

One obtains 
\bea
-W_{eff}^{\Delta B\neq =0} = L_{ABCD} ({1\over 2}\epsilon Q_A Q_B Q_C L_D) +
                          R_{ABCD}  (\epsilon {\bar e}_A {\bar u}_B {\bar u}_C {\bar d}_D)
\eea

where the coeffcients are 

\bea
L_{ABCD} =  {\cal S}_1^1  h_{AB} h_{CD} + {\cal S}_1^2  f_{AB} h_{CD} +
 {\cal S}_2^1  f_{AB} h_{CD} + {\cal S}_2^2  f_{AB} f_{CD} \eea

and 
\bea
R_{ABCD} = L_{ABCD} -i{\sqrt 2} {\cal S}_1 ^4 f_{AB} h_{CD} 
-i {\sqrt 2} {\cal S}_2 ^4 f_{AB} f_{CD} 
\eea

here 
\bea 
h_{AB} = 2 {\sqrt 2} h_{AB}' \qquad  f_{AB} = 4 {\sqrt 2} f_{AB}' 
\eea

We note that this expression and the "Clebsches" contained in it , 
as well as the new baryon decay "channel"  mediated by the triplets contained in 
${\bf{\overline{\Sigma}_{126}(10,1,3)}}$ (i.e $t^{(4)} $ ), 
( the same PS multiplet that contains the Higgs field responsible 
for the right handed neutrino Majorana mass) 
have not, to our knowledge, appeared  previously in the literature.
Previous work \cite{babpatwil} on ${\bf{\overline{\Sigma}_{126}}}$ mediated 
decay focussed only on the multiplets $t^{(2)},{\bar t}^{(2)} $ and 
found  that there was no  contribution of $t^{(4)},{\bar t}^{(4)}$ in their models. 
This new   channel nominally  strengthens the emergent link between
 neutrino mass and baryon decay. 
 Note however that $t^{(4)}$ couples only to the RR combinations 
$( {\bar d}_A {\bar\nu}_B + {\bar e}_A {\bar u}_B ) $ and as such its exchange 
 will contribute only to the RRRR channel which , at least in SO(10), seems
\cite{babpatwil}  generically suppressed except at very large $tan \beta$.
 However the mixing in the 
triplet mass matrix could also strengthen the effects of this channel.
 From this expression together with information on the ${\bf{10,{\overline{126}}}}$ 
content of the light doublets, the baryon  decay rates can be calulated
following a by now standard proceedure\cite{bdec}.
The couplings $h_{AB} , f_{AB}$ are tightly constrained\cite{seesaw1}
 by the fit of fermion masses . Thus the 
the number of free parameters is relatively low  and this will allow a fairly restrictive 
estimate of these processes in this model. Details will be given elsewhere . 

 {\section{Discussion}} In this paper we have
carried out the tedious calculations required to provide a tool
kit for ready translation of any SO(10) invariant one is likely to
encounter in the course of SO(10) GUT model building into a
convenient form where the fields carry unitary group labels. This
allows calculation of all ``Clebsch-Gordon" coefficients relevant to
SO(10) GUT models : including many which were as yet unavailable in the literature.
In addition we have obtained a very explicit description of the
action of D parity on all fields. This allows one to follow the
operation of D-parity, which implements Left-Right symmetry i.e.
parity, in such theories. This translation is necessary in order
to carry out RG analysis based on calculated mass spectra and will
also be useful to obtain more accurate estimates of threshold
uncertainties.

 We used the previously unavailable "Clebsches" to calculate 
the Mass matrices of the
 doublets and triplets that mix with with those contained in the 
$\bf{10,{\overline {126}}}$ multiplets. We also calculated the clebsches for the 
couplings of the 
doublets and triplets  contained in the 
$\bf{10,{\overline {126}}}$  to light matter supermultiplets contained in the spinorial
$\bf{16} $. These allowed us to obtain the 
crucial bare  effective superpotential for Baryon  number violation 
in this``minimal Supersymmetric GUT" which was proposed as long back as 1982 \cite{aulmoh,ckn}
but for which these quantities were hitherto unavailable. Indeed, some very recently published 
expressions \cite{fukujanac}, are erroneous not only in the values of numerical coeffcients 
but even in the channels (they have an anti-triplet from 
$({\overline{10}},1,3)\in {\overline {126}}$ 
coupling to $QL$ : but 16 x 16 = 126 +... , implies that ${\overline {126}}$ contains 
$(10,1,3)$ {\it{not}} $({\overline{10}},1,3)$ !!). 
 In view of the topicality and 
phenomenological success of such  GUTs \cite{seesaw1}  along with the tight 
experimental constraints on most of its (non-soft) parameters these  results 
 may prove  of general 
interest in the GUT community.

 Furthermore since  our method 
reduces all the difficulties of reducing Spin(10) invariants 
to a standard manipulation of Unitary group labels it may 
find appeal to those who would like to eschew the use of a computer to calculate 
the coupling coeffcients ( where that is even feasible !). 
 A systematic study of
related theories along the lines of the program outlined in
\cite{abmsv,geneal} using the tools developed here will be
presented elsewhere. We hope that our
techniques and results will be found useful by other practitioners
of the unwieldy and  - so far - somewhat obscure art of SO(10) GUT
building, even if only due to the simple minded and (perhaps)
objectionably explicit approach we have taken to the analysis of
this niggling group theoretical problem. Our rules may also be
applied in other contexts where one encounters these groups for
example in 10 dimensional field theories where the Lorentz group
is SO(1,9) and a translation to SU(4) labels instead of
SO(6) labels for the compactified sector may prove more
convenient, specially for spinorial indices.

\centerline{\bf{Acknowledgements}}CSA is grateful to Goran  Senjanovic for
discussions and encouragement and for hospitality at ICTP, Trieste
where this work was initiated. We also thank Borut Bajc and
Francesco Vissani for discussions and collaboration
and Alejandra Melfo in particular for numerous collaborative discussions as well as 
for pointing out certain errors in the coeffcients of eqn.(141) (142).
 This work is supported by the
Department of Science and Technology, Government of India under
project No.SP/S2/K-07/99.

\centerline{\bf{APPENDIX}} In this section we have collected
useful $SO(6)\leftrightarrow SU(4)$, $SO(4)\leftrightarrow
SU(2)\times SU(2)$ identities for the convenience of the reader
while translating invariants of his choice using our methods.
\setcounter{equation}{0}
\renewcommand{\theequation}{A.\arabic{equation}}

\subsection{SO(6)}
Two vectors :
\begin{equation}
V_{a}W_{a}={1 \over{2}}\widetilde{V}^{\mu\nu}W_{\mu\nu}
\qquad\qquad \widetilde{V}^{\mu\nu}\equiv {1 \over{2}}\epsilon^{\mu\nu\lambda\sigma}V_{\lambda
\sigma}
\ee
The ``raised" versions of eqn. (\ref{AD}),(\ref{TPD}),(\ref{TMD})
are \bea
A^{\mu\nu,\lambda\sigma}&=&+A_{\theta}^{~[\mu}\epsilon^{\nu]\lambda\sigma\theta}\\
T_{(+)}^{\mu\nu,\lambda\sigma,\theta\delta}&=&-\epsilon^{\mu\nu\gamma[\theta}\epsilon^{\delta]\lambda\sigma\omega}T_{\gamma\omega}\\
T_{(-)}^{\mu\nu,\lambda\sigma,\theta\delta}&=&T^{[\mu_{[}\lambda}\epsilon^{\nu^{]}\sigma_{]}\theta\delta}
\eea Two index antisymmetric tensors  :\be
A_{ab}B_{ba}=2{A_{\nu}}^{\mu}{B_{\mu}}^{\nu} \ee
Two index  traceless symmetric tensors
\be
\widehat{S}_{ab}\widehat{R}_{ba}={1 \over{4}}\widehat{S}^{\mu\nu,\lambda\sigma}\widehat{R}
_{\mu\nu,\lambda\sigma}
\ee
Three index antisymmetric tensors :
\be
T_{abc}U_{abc}={1 \over{2}}{(T^{+}_{abc}U^{-}_{abc}+T^{-}_{abc}U^{+}_{abc})}
=3(T_{\mu\nu}\overline{U}^{\mu\nu}+\overline{T}^{\mu\nu}U_{\mu\nu})
\ee
where  $T^{+}_{abc},T^{-}_{abc}$ are self-dual and anti-self-dual parts of
$T_{abc}$.\\
%
%
%
Mixed two index and three index antisymmetric tensors :
\bea
A_{ab}T^{+}_{acd}U^{-}_{bcd}&=&-4({A_{\nu}}^{\mu}T_{\mu\lambda}\overline{U}^
{\nu\lambda})\\
\phi_{abcd}T^{+}_{abe}U^{-}_{cde}&=&8i\phi_{\nu}^{~\mu}T_{\mu\lambda}\overline{U}^{\nu\lambda}\\
\epsilon_{abcdef}A_{ab}T^{+}_{cdg}U^{-}_{efg}&=&16i({A_{\nu}}^{\mu}T_{\mu
\lambda}\overline{U}^{\nu\lambda}) \eea Three two index
antisymmetric  tensors \bea A_{ab}B_{bc}C_{ca}&=&-trA[B,C]\\
\epsilon_{abcdef}A_{ab}B_{cd}C_{ef}&=&-8itrA{\{B,C}\} \eea
Three two index symmetric traceless tensors :
\be
\widehat{S}_{ab}\widehat{R}_{bc}\widehat{T}_{ca}={1 \over{8}}\widehat{S}^{\mu\nu,\lambda\sigma}
\widehat{R}_{\lambda\sigma}
^{\theta\delta}\widehat{T}_{\theta\delta,\mu\nu}
\ee
Two vectors and two index tensors : \bea Antisymmetric~~
V_{a}W_{b}A_{ab}=V_{\mu\nu}W^{\nu\lambda}{A_{\lambda}}^{\mu}\\
Symmetric~ traceless~~ V_{a}W_{b}\widehat{S}_{ab}={1
\over{4}}\widetilde{V}^{\mu\nu}\widetilde{W}^{\lambda
\sigma}\widehat{S}_{\mu\nu,\lambda\sigma} \eea
Vector with two index and three index antisymmetric tensors : \bea
V_{a}A_{bc}T_{abc}=V_{a}A_{bc}{(T^{+}_{abc}+T^{-}_{abc})\over\sqrt{2}}&=&{\sqrt{2}}
{({{-\tilde{V}}}^{\mu\nu}{A_{\nu}}^{\lambda}T_{\mu\lambda} +
{V}_{\mu\nu}{A_{\lambda}}^{\nu}\overline{T}^{\mu\lambda})}\\
\epsilon_{abcdef}V_{a}A_{bc}T_{def}&=&(i3!{\sqrt{2}}){({{\tilde{V}}}^{\mu\nu}{A_{\nu}}^
{\lambda}T_{\mu\lambda} +
{V}_{\mu\nu}{A_{\lambda}}^{\nu}\overline{T}^{\mu \lambda})} \eea
Two antisymmetric (A,B) and one symmetric traceless (S) two index tensor :
\be
A_{ab}\widehat{S}_{bc}B_{ca}= {1\over 2} A^{\mu}_{[\nu} \widehat{S}_{\lambda ]\mu}^{\lambda\delta}
 B_{\delta}^{\nu}
\ee
For the product of a tensor with two antisymmetric indices and two tensors
with two symmetric indices :
\be
A_{ab}\widehat{S}_{bc}\widehat{R}_{ca}={1
\over{2}}{A_{\kappa}}^{\mu}\widehat{S}^{\nu\kappa,\lambda\sigma}
\widehat{R}_{\lambda\sigma,\mu\nu} \ee
%
%
%
%
%
%
%
\subsection{SO(6) Invariants with Spinors}
For SO(6) sector $C \equiv C_{2}^{(3)}$
\begin{eqnarray}
\psi^{T}C\gamma_{a}\chi{V}_{a}&=&\sqrt{2}(\widehat{\psi}^{\mu}\widehat{\chi}^{\nu}V_{\mu\nu}-\psi_{\mu}\chi_{\nu}\widetilde{V}
^{\mu\nu})\\
\psi^{T}C\gamma_{a}\gamma_{b}\chi{V}_{a}W_{b}&=& 2(\widehat{\psi}^{
\mu}\chi_{\nu}V_{\mu\lambda}\widetilde{W}^{\nu\lambda}+
\psi_{\mu}\widehat{\chi}^{\lambda}\widetilde{V}^{\mu\nu}W_{\lambda\nu})\\
\psi^{T}C\gamma_{a}\gamma_{b}\chi{A}_{ab}&=&4{A_{\nu}}^{\mu}
(-\psi_{\mu}\widehat\chi^{\nu}+\widehat\psi^{\nu}\chi_{\mu})\\
\psi^{T}C\gamma_{a}\gamma_{b}\gamma_{c}\chi{T}_{abc}&=&12(\overline{T}
^{\mu\nu}\psi_{\mu}\chi_{\nu}-T_{\mu\nu}\widehat{\psi}^{\mu}\widehat\chi^{\nu})\\
\psi^{T}C\gamma_{a}\gamma_{b}\gamma_c \chi{V}_{a}W_{b} U_c &=&
2 {\sqrt 2} (\psi_{\mu} \chi_{\delta}{\tilde V}^{\mu\nu} W_{\nu\theta}U^{\theta\delta}
-{\widehat \psi}^{\mu} {\widehat \chi}^{\nu} V_{\mu\theta}
 W_{\nu\delta}\tilde{U}^{\theta\delta})\\
\psi^{T}C\gamma_{a}\gamma_{b}\gamma_{c}\chi{V}_{a}A_{bc}&=&{-2\sqrt{2}}
(\widehat{\psi}^{\mu}\widehat\chi^{\nu} V_{\mu\lambda}
+\psi_{\mu}\chi_{\lambda} \widetilde{V}^{\mu\nu})
{A_{\nu}^{\lambda}}
\end{eqnarray}
 
\subsection{SO(4)}
 Two vectors :
\be
V_{\tilde\alpha}W_{\tilde\alpha}=-V^{\alpha\dot\alpha}W_{\alpha\dot\alpha}
\ee
Two antisymmetric tensors :
\be
A_{\tilde\alpha\tilde\beta}B_{\tilde\alpha\tilde\beta}=2(\vec{A}_{R}.\vec{B}
_{R}+\vec{A}_{L}.\vec{B}_{L})
~~~,~~~
\epsilon_{\tilde\alpha\tilde\beta\tilde\gamma\tilde\delta}A_{\tilde\alpha
\tilde\beta}B_{\tilde\gamma\tilde\delta}=4(\vec{A}_{R}.\vec{B}
_{R}-\vec{A}_{L}.\vec{B}_{L})
\ee

Three antisymmetric tensors :
\bea
A_{\tilde\alpha\tilde\beta}B_{\tilde\beta\tilde\gamma}
C_{\tilde\gamma\tilde\alpha}&=&
-{1 \over \sqrt{2}}{\{A_{(R)}}^{\dot\alpha\dot\beta}{B_{(R)~\dot
\beta}}^{\dot\gamma}C_{(R)~\dot\gamma\dot\alpha}\nonumber
+A_{(L)}^{\alpha\beta}{B_{(L)~
\beta}}^{\gamma}C_{(L)~\gamma\alpha}\}\nonumber \\
&=& {\sqrt{2}}\{  {\vec A}_R \cdot ({\vec B}_R \times {\vec C}_R)
+  {\vec A}_L \cdot ({\vec B}_L \times {\vec C}_L)\}
\eea

Two vectors and  an antisymmetric tensor :
\be
V_{\tilde\alpha}W_{\tilde\beta}A_{\tilde\alpha\tilde\beta}={1 \over \sqrt{2}}
{\{V}^{\alpha\dot\alpha}{W_{\alpha}}^{\dot\beta}
{A_{(R)~\dot\alpha\dot\beta}}+
V^{\alpha\dot\alpha}{W^{\beta}}_{\dot\alpha}A^{(L)}_{\alpha\beta}\}
\ee
When the indices are contracted with the invariant tensor of SO(4) :
\be
\epsilon_{\tilde\alpha\tilde\beta\tilde\gamma\tilde\delta}V_{\tilde\alpha}W_
{\tilde\beta}A_{\tilde\gamma\tilde\delta}=\sqrt{2}{\{V}^{\alpha\dot\alpha}{W_
{\alpha}}^{\dot\beta}{A_{{(R)}~\dot\alpha\dot\beta}}-
V^{\alpha\dot\alpha}{W^{\beta}}_{\dot\alpha}A^{(L)}_{\alpha\beta}\}
\ee
Two traceless symmetric tensors :
\be
\widehat{S}_{\tilde\alpha\tilde\beta}\widehat{R}_{\tilde\alpha\tilde\beta}=\widehat{S}
^{\alpha\beta,\dot
\alpha\dot\beta}\widehat{R}_{\alpha\beta,\dot\alpha\dot\beta}
\ee
Three symmetric tensors :
\be
\widehat{S}_{\tilde\alpha\tilde\beta}\widehat{R}_{\tilde\beta\tilde\gamma}\widehat{T}
_{\tilde\gamma\tilde
\alpha}=- \widehat{S}^{\alpha\beta,\dot\alpha\dot\beta}\widehat{R}_{\beta\gamma,\dot\beta\dot
\gamma}{{\widehat{T}^{\gamma}}_{\alpha},^{\dot\gamma}}_{\dot\alpha}
\ee
Two vectors and a symmetric tensor :
\be
V_{\tilde\alpha}W_{\tilde\beta}\widehat{S}_{\tilde\alpha\tilde\beta}=V^{\alpha\dot
\alpha}W^{\beta\dot\beta}\widehat{S}_{\alpha\beta,\dot\alpha \dot\beta}
\ee
Two antisymmetric and one symmetric tensor :
\be
A_{\tilde\alpha\tilde\beta}B_{\tilde\beta\tilde\gamma}\widehat{S}_{\tilde\gamma\tilde
\alpha}=-{1 \over 2}{\{A}_{(R)}^{\dot\alpha\dot\beta}B_{(L)}^{\alpha\beta}+
A_{(L)}^{\alpha\beta}B_{(R)}^{\dot\alpha\dot\beta}\}\widehat{S}_{\alpha\beta,\dot\alpha
\dot\beta}
\ee
 One antisymmetric and  two symmetric :
\be
A_{\tilde\alpha\tilde\beta}\widehat{S}_{\tilde\beta\tilde\gamma}
\widehat{R}_{\tilde\gamma\tilde\alpha} =
-{1 \over {\sqrt 2}}
( {A}^{(R)}_{\dot\alpha\dot\beta}
 \widehat{S}_{\alpha}^{\gamma\dot\beta\dot\gamma} +
{A}^{(L)}_{\alpha\beta}\widehat{S}^{\gamma\beta\dot\gamma}_{\dot\alpha})
\widehat{R}_{\gamma\dot\gamma}^{\alpha\dot\alpha}
\ee
%
%
%
%
%
%
%
%
%
%
%
%
\subsection{SO(4) Invariants with Spinors}
For the SO(4) sector $C \equiv C_{2}^{(2)}$
\begin{eqnarray}
\psi^{T}C\gamma_{\tilde\alpha}\chi{V}_{\tilde\alpha}&=&{\sqrt{2}}(\psi_{\alpha}\chi
_{\dot\alpha}-{\psi}_{\dot\alpha}\chi_{\alpha})V^{\alpha\dot\alpha}\\
\psi^{T}C\gamma_{{\tilde\alpha}}\gamma_{{\tilde\beta}}\chi{V}
_{\tilde\alpha}W_{\tilde\beta}&=
&   2{\psi}_{\alpha}\chi_{\beta} V^{\alpha \dot\alpha}{W}^{\beta}
_{\dot\alpha}- 2{\psi}_{\dot\alpha}\chi_{\dot\beta}
V^{\alpha\dot\alpha} W_{\alpha}^{\dot\beta} \\
\psi^{T}
{{C^{(2)}_{2}}\brace C^{(2)}_{1}}
\gamma_{\ta}\gamma_{\tb}
\chi{A}_{\ta\tb}
&=& -2\sqrt{2}\{ A^{\dot\alpha\dot\beta}
\psi_{\dot\alpha}
\chi_{\dot\beta}
\mp A^{\alpha\beta}
\psi_{\alpha}\chi_{\beta}\}
\label{so4}
\end{eqnarray}

\end{document}